\documentclass[sigconf]{acmart}

\usepackage{tikz}
\usepackage{amsmath}

\usepackage{wrapfig}
\usepackage{comment}
\usepackage{array}
\usepackage{hyperref}
\usepackage{tablefootnote}
\usepackage{cleveref}
\usepackage{xcolor}

\hypersetup{
  colorlinks,
  linkcolor={green!80!black},
  citecolor={red!70!black},
  urlcolor={blue!70!black}
}
\usepackage{multirow}
\usepackage{booktabs}
\usepackage{subcaption}
\usepackage{caption}
\usepackage{float}
\usepackage{enumerate}
\usepackage{adjustbox}

\usepackage{eurosym}
\usepackage{graphicx,multicol}
\usepackage{xspace}

\usepackage{balance}
\usepackage{epsfig}
\usepackage{enumitem}
\definecolor{blue-green}{rgb}{0.0, 0.87, 0.87}
\definecolor{bleudefrance}{rgb}{0.19, 0.55, 0.91}

\usepackage[utf8]{inputenc}
\usepackage[T1]{fontenc}

\newcommand{\jg}[1]{{\color{black} #1}}

\newcommand{\CamR}[1]{{\color{black} #1}}
\newcommand{\CamRd}[1]{{\color{black} #1}}
\newcommand{\GramR}[1]{{\color{black} #1}}

\copyrightyear{2021} 
\acmYear{2021} 
\setcopyright{acmcopyright}\acmConference[IMC '21]{ACM Internet Measurement Conference}{November 2--4, 2021}{Virtual Event, USA}
\acmBooktitle{ACM Internet Measurement Conference (IMC '21), November 2--4, 2021, Virtual Event, USA}
\acmPrice{15.00}
\acmDOI{10.1145/3487552.3487861}
\acmISBN{978-1-4503-9129-0/21/11}

\begin{document}
\title[Formulation and Evidence of Facebook Nanotargeting with non-PII]{Unique on Facebook: Formulation and Evidence of (Nano)targeting Individual Users with non-PII Data}

\author{José González-Cabañas}\orcid{0000-0003-2055-4433}
\affiliation{%
  \institution{
  Universidad Carlos de III Madrid}
}
\email{jgcabana@it.uc3m.es}

\author{Ángel Cuevas}\orcid{0000-0002-5738-0820}
\affiliation{%
  \institution{
  Universidad Carlos III de Madrid\\ 
  UC3M-Santander Big Data Institute}
}
\email{acrumin@it.uc3m.es}

\author{Rubén Cuevas}\orcid{0000-0002-1440-8360}
\affiliation{%
  \institution{
  Universidad Carlos III de Madrid\\ 
  UC3M-Santander Big Data Institute}
}
\email{rcuevas@it.uc3m.es}

\author{Juan López-Fernández}
\affiliation{%
  \institution{GTD System \& Software Engineering}
}
\email{juan.lopez@gtd.eu}

\author{David García}\orcid{0000-0002-2820-9151}
\affiliation{%
  \institution{Graz University of Technology}
}
\email{dgarcia@tugraz.at}

% The default list of authors is too long for headers
\renewcommand{\shortauthors}{J. González-Cabañas et al.}

\begin{abstract}
The privacy of an individual is bounded by the ability of a third party to reveal their identity. Certain data items such as a passport ID or a mobile phone number may be used to uniquely identify a person. These are referred to as Personal Identifiable Information (PII) items. Previous literature has also reported that, in datasets including millions of users, a combination of several non-PII items (which alone are not enough to identify an individual) can uniquely identify an individual within the dataset. 
In this paper, we define a data-driven model to quantify the number of interests from a user that make them unique on Facebook. To the best of our knowledge, this represents the first study of individuals' uniqueness at the world population scale. Besides, users' interests are actionable non-PII items that can be used to define ad campaigns and deliver tailored ads to Facebook users. We run an experiment through 21 Facebook ad campaigns that target three of the authors of this paper to prove that, if an advertiser knows enough interests from a user, the Facebook Advertising Platform can be systematically exploited to deliver ads exclusively to a specific user. We refer to this practice as \textit{nanotargeting}. Finally, we discuss the harmful risks associated with nanotargeting such as psychological persuasion, user manipulation, or blackmailing, and provide easily implementable countermeasures to preclude attacks based on nanotargeting campaigns on Facebook.
\end{abstract}




%\keywords{Nanotargeting; Facebook; Privacy; Non-PII; Online Advertising}
\maketitle
\section{Introduction}

In the current hyper-connected world, an individual's privacy is bounded by the amount of information a third party needs to know to identify them. Beyond Personal Identifiable Information (PII), e.g., email address, phone number, postal address, passport ID, etc., which by definition uniquely identifies an individual, a user could be also uniquely identified by the combination of a certain number of non-PII elements. Defining the number of non-PII items required to uniquely identify a user is of paramount importance to understand the actual limits of users' privacy. Preliminary studies in the area of users' uniqueness have demonstrated that the spatio-temporal information of 4 mobile phone calls \cite{de2013unique} or 4 credit card purchases \cite{de2015unique} uniquely identify more than 90\% of the users in a dataset with 1.5 million people. Similarly, 3 demographic items (gender, ZIP code, and birth date) are \GramR{enough to identify 63\% of US citizens} within the US 2000 census \cite{Golle:2006:RUS:1179601.1179615}. However, these studies are limited either to a small user base or to a single country. 

In this paper, we present, to the best of our knowledge, the first study that addresses individuals' uniqueness considering a user base at the worldwide population's order of magnitude. The focus of our study is Facebook (FB), a platform having more than 2.8B active users \cite{fb_q4_2020} at the end of 2020. The non-PII items that we consider in our analysis are the interests that FB assigns to users based on their online and offline activity. Users' interests represent a very important asset for FB since its revenue model is based on delivering relevant ads to users. Many advertisers use the FB advertising platform to create ad campaigns to reach users based on their interests. 

The first contribution of this paper is a data-driven model that provides the metric $N_P$, which is defined as the number of interests that uniquely identify a user on FB with a probability $P$. For instance, $N_{50}=12$ means that the probability to uniquely identify a user with 12 interests is 50\%. To obtain $N_P$, we study the audience size for thousands of FB \GramR{audiences formed by combining between} 1 and 25 interests. We retrieve the size of the audiences used in our study from the FB Ads Campaign Manager\cite{FBadsmanager}. Moreover, to create the combinations of interests we use real interest sets from 2.4k FB users that installed \CamR{the FDVT (Data Valuation Tool for Facebook Users) \cite{FDVT}} browser extension that collects the interests FB has assigned them.%\footnote{We do not provide the actual name of the referred browser extension for double-blind purposes. Furthermore, we have deleted the name of the browser extension in all figures throughout the paper where it was present.} %The users who installed the browser extension permitted us to use the collected information for research purposes.

The results from our model reveal that the 4 rarest interests or 22 random interests from the interests set FB assigns to a user make them unique on FB with a 90\% probability.

In contrast to the non-PII items considered in some of the previous works studying uniqueness (e.g., credit card transactions or mobile phone calls), users' interests on FB are intentionally designed to be actionable through FB ad campaigns. Therefore, since a user can be uniquely identified by a set of interests on FB, it may be possible to configure \GramR{a Facebook} ad campaign that reaches exclusively a single user. We refer to this practice as \emph{nanotargeting} in this paper. 

Nanotargeting is a potentially harmful practice. The literature from the \textit{psychological persuasion} discipline has demonstrated that persuading an individual is easier if you can create tailored messages to the psychological characteristics and motivations of that person \cite{Kosinski_PNAS}. In this context, nanotargeting might be a very powerful tool for attackers willing to manipulate a specific individual. Nanotargeting could be also used to blackmail users.

\jg{To our knowledge, there is no previous evidence of the possibility of systematically exploiting the FB advertising platform to implement nanotargeting based on non-PII data.} 

The second contribution consists in providing the first empirical evidence that nanotargeting can be systematically implemented on FB by just knowing a random set of interests of the targeted user. In particular, we have configured nanotargeting ad campaigns targeting three authors of this paper. \jg{We tested the results of our data-driven model by creating tailored audiences for each targeted author using combinations of 5, 7, 9, 12, 18, 20, and 22 randomly selected interests from the list of interests FB had assigned them. In total, \GramR{we ran 21 ad campaigns between October and November 2020 to demonstrate} that nanotargeting is feasible today.}

Our experiment validates the results of our model, showing that if an attacker knows 18+ random interests from a user, they will be able to nanotarget them with a very high probability. In particular, 8 out of the 9 ad campaigns that used 18+ interests in our experiment successfully \GramR{nanotargeted the chosen user}.

After proving the plausibility to systematically conduct nanotargeting ad campaigns on FB nowadays, the last contribution of this paper is focused on discussing and proposing solutions to protect users from the potential pernicious risks associated with nanotargeting (e.g., manipulation, blackmailing, etc.). First, we add new functionality to \CamR{the FDVT} browser extension to reveal to users what interests are more harmful to their privacy (i.e., those associated with a smaller audience size) using a simple color scale. Our solution also enables users to delete those interests with a single click. Second, we propose easily implementable measures that FB could adopt to preclude nanotargeting attacks through its advertising platform.

\section{Background}
\label{sec:background}

In this section, we describe the technological venues that we use in this paper. Particularly, we first describe the FB advertising platform which serves a twofold purpose in our research. We use it to (i) retrieve the audience sizes that serve as input to our model of users' uniqueness on FB, and (ii) configure the advertising campaigns of our nanotargeting experiment. Second, we describe the \CamR{FDVT} browser extension, which provides the set of interests that we use as input in our analysis of users' uniqueness on FB.

\subsection{Facebook Ad Platform Overview}

Advertisers configure their ad campaigns on FB through the \emph{FB Ads Campaign Manager} \cite{FBadsmanager}. This is a dashboard where advertisers define the audience (\textit{i.e.}, user profile) they want to target. 
The FB Ads Campaign Manager offers advertisers a wide range of configuration parameters such as (but not limited to): \emph{location} (country, region, etc.), \emph{demographic parameters} (gender, age, etc.), \emph{behaviors} (mobile device, OS and/or web browser used, etc.), and \emph{interests} (sports, food, etc.). In principle, all these attributes are considered non-PII data since they cannot be used alone to identify a user. Moreover, the FB Ads \GramR{Campaign Manager conveys the size} of the audience configured in the dashboard through the so-called \emph{Potential Reach} parameter. This parameter reports the number of Monthly Active Users (MAU) on FB matching the defined audience, which by definition is the audience size. In addition to the dashboard, the FB Ads Campaign Manager offers advertisers an API to automatically retrieve the Potential Reach for any audience. We leveraged that API to retrieve the \textit{Potential Reach} associated with the audiences used to build our model.

On the other hand, FB assigns to each user a set of interests, referred to as \emph{ad preferences}. The ad preferences of a user are inferred from the data and activity of the user on FB and other websites and online services where FB is present. These ad preferences are indeed the interests offered to advertisers in the FB Ads Manager. Therefore, if a user is assigned \textit{``Italian food''} within their list of interests, they will be a potential target of any FB advertising campaign configured to reach users interested in \textit{``Italian food''}. It is important to note that interests in the FB ad ecosystem are global, thus there are not specific interests per country. 

The only compulsory parameter to define an audience in FB is the location. An advertiser can combine that location with any of the other available attributes. Hence, we can obtain the number of FB users in a particular location (e.g., zip code, town, country, etc) or group of locations that have been assigned a particular interest (or group of interests). Due to privacy reasons, the minimum \emph{Potential Reach} value that FB returns for any audience since 2018 is 1000. Previously, this limit was only 20. In this paper, we use a dataset collected in January 2017, so that our data is bounded by an audience size limitation of 20 users. %Note that, the 1000 limit in the \emph{Potential Reach} only affects the analysis of audience sizes but not the potential to nanotarget individuals. 

\CamR{The \emph{Potential Reach} limitation does not prevent advertisers running campaigns for audiences including less than 1000 users, but it only prevents the advertiser from knowing the actual size of the targeted audience. Even more, a previous work \cite{fbto100} describes a mechanism to reduce under the current Facebook limits the minimum \emph{Potential Reach} from 1000 to 100. This method may help advertisers to infer the size of their targeted audience even if this is lower than 1000. Therefore, any author willing to replicate our uniqueness analysis would be limited by a threshold of 100 instead of 20.}

It is important to note that, at the time we collected the dataset, the FB Ads Manager had two limitations. First, we could not create queries including more than 25 interests (this limitation remains nowadays). Second, the FB Ads Manager did not include the whole world as a possible location (this option is available nowadays). Instead, it requested to introduce a specific location (country, region, town, ZIP code, etc.) or group of locations. The maximum number of locations allowed in a query was 50. Therefore, to maximize the number of users addressed in our research, we ran our queries for a location set including the 50 countries with the largest number of FB users (see Appendix \ref{ap:tabla50}). These countries accounted for 1.5B active users, which corresponded to 81\% of the overall FB when we collected the data \cite{fb_end_report_2016}.

%The FB Ads Manager provides detailed information about the configured audience. The most relevant element for our paper is the \emph{Potential Reach} that reports the number of monthly active users (MAU) on FB matching the defined audience. 
%The number of users assigned to one interest go from this minimum of 20 to extremely large audiences, for example, the interest \textit{Technology} is assigned to 1.66 Billion users. 

Finally, Facebook also allows advertisers to target users based on PII data items through the \emph{Custom Audience} \cite{custom_audiences_fb} functionality on its advertising platform. A custom audience refers to a list of users identified by a PII item (e.g., mobile phone number, email address, etc.). A Facebook ad campaign based on a custom audience has the goal of reaching the users included in such custom audience list. To this end, FB finds the registered users who match any of the PII items included. FB imposes two important requirements for the use of a custom audience: $(i)$ Advertisers are responsible for obtaining explicit consent from the users included in the audience to be targeted with FB Custom Audience advertising campaigns. Failing to do so may imply the advertiser/attacker is breaking personal data regulations such as the General Data Protection Regulation (GDPR) \cite{GDPR} in Europe. This requirement appears not to be needed when using non-PII attributes; $(ii)$ The minimum number of users forming a custom audience has to be 100. 
Although custom audiences are of high interest in the context of privacy studies they require PII data and thus are out of the scope of this paper.

\subsection{FDVT Browser Extension}

The list of interests considered in this work is obtained from 2,390 real FB users that installed \CamR{the FDVT} browser extension \cite{FDVT} before January 2017. \jg{Our extension was publicly released in Oct. 2016 and our user base is formed by users that freely decided to install it.} The main functionality of \CamR{the FDVT} browser extension is to provide users with a real-time estimation of the revenue they generate for FB out of the ads they receive while browsing on FB. To compute the revenue estimation the users provide in a registration process a few demographic parameters that allow us to build an audience from them: (i) Country of residence (compulsory), (ii) Gender (optional), (iii) Age (optional), and (iv) Relationship status (optional). The \CamR{FDVT} browser extension collects (among other data) the interests FB assigns to the user. To this end, the browser extension parses the user's ad preferences page \cite{FBadsprefs}. %\footnote{\url{https://www.facebook.com/adpreferences/ad_settings}}. 

\subsection{Ethics (IRB and Users consent)}

Our research is committed to complying with ethical and legal standards. From the legal point of view, we are subject to the General Data Protection Regulation (GDPR) \cite{GDPR} that applies to all EU countries. To comply with the GDPR, at the time of installing the \CamR{FDVT} browser extension (i.e., registration process), all users have to: (i) proactively accept (opt-in) \CamR{the Terms of Use \cite{fdvt_terms_of_use} and Privacy Policy \cite{fdvt_policy_privacy}}%\footnote{To comply with the double-blind process we do not reveal these links. If accepted, the final version of the manuscript will include them}
, and (ii) provide us with explicit permission (opt-in) to use the information anonymously collected for research purposes. From the ethical point of view, the ethics committee of the authors' institution provided an IRB approval to develop the \CamR{FDVT} browser extension as a part of an H2020 European project and the research activities derived from it.

\CamR{In Section 5, we run real ad campaigns aiming at nanotargeting 3 of the authors of this paper using a different number of interests. In some cases, these campaigns also reach other users different from the targeted authors. We have used ads (see Figure \ref{fig:ad_ap1}) that simply promote \CamR{the FDVT browser} extension in all our ad campaigns. In addition, our ads do not track ad impressions. Therefore, there is no way we can obtain the identity of the users receiving the ad. In case some user clicks on the ad, they are forwarded to the FDVT's website \cite{fdvt_url} (the website project of our FDVT plug-in). As in most standard web services we collected the IP address of the device that has opened the connection. This is a common practice for multiple reasons from which the most important one is security. To further protect users' privacy, in the context of this experiment, we have pseudonymized the IP address using a (secret) hash-function and key. In addition, the only users targeted in the ad campaigns of Section \ref{sec:nanotargeting_experiment} are authors of the paper who are aware and accept the purpose of the nanotargeting experiments.}

\section{Dataset}
\label{sec:dataset}
Our dataset is created from 2,390 real users that installed our \CamR{FDVT} browser extension between October 2016 (public release) and January 2017. Of these users, 1,949 declared to be men, 347 to be women, and 94 did not disclose their gender. Furthermore, following the age group classification proposed in \cite{erikson1998life}, 117 users are adolescents (aged 13-19), 1374 early adults (aged 20-39), 578 adults (aged 40-64), 19 matures (aged 65+), and 302 did not provide their age. Finally, our 2,390 users stated to be located in 80 different countries. A detailed breakdown of the number of users per country in our dataset can be found in Appendix \ref{ap:distribution}.

\begin{figure}[!t]
\centering
	\includegraphics[width=1\columnwidth]{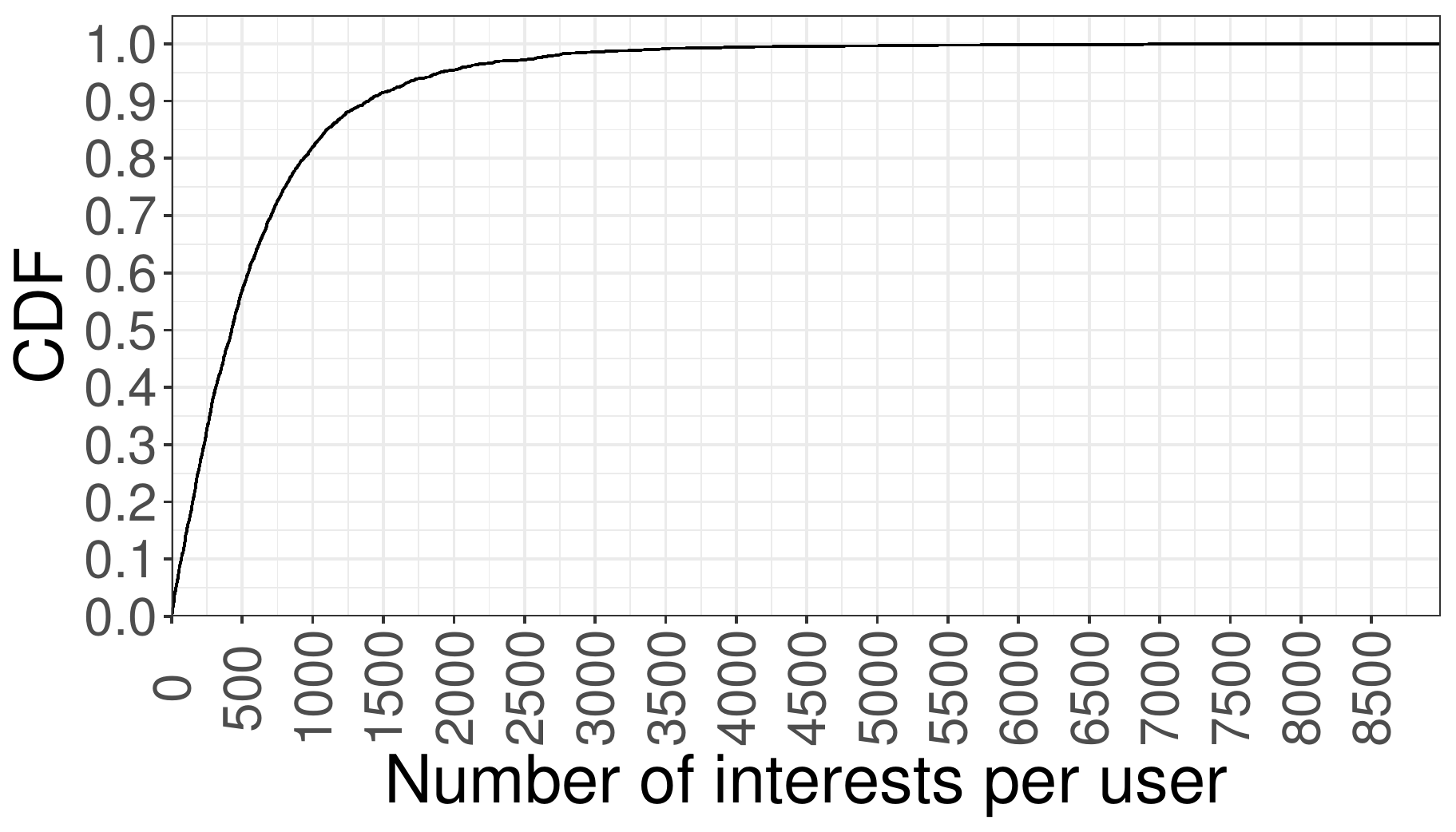}\hfill
	%\vspace{-0.2cm}
	\caption{CDF showing the distribution of the number of interests assigned to the 2,390 users of our dataset.}
	\label{fig:cdf_user_interests}
  \hfill
\end{figure}

We retrieved 1.5M occurrences out of 99k unique FB interests assigned to the 2,390 users. Figure \ref{fig:cdf_user_interests} displays the CDF of the number of interests per user. The number of interests FB assigned to an individual user in our dataset ranges between 1 and 8,950, with a median of 426 interests. 

\begin{figure}[!t]
\centering
	\includegraphics[width=1\columnwidth]{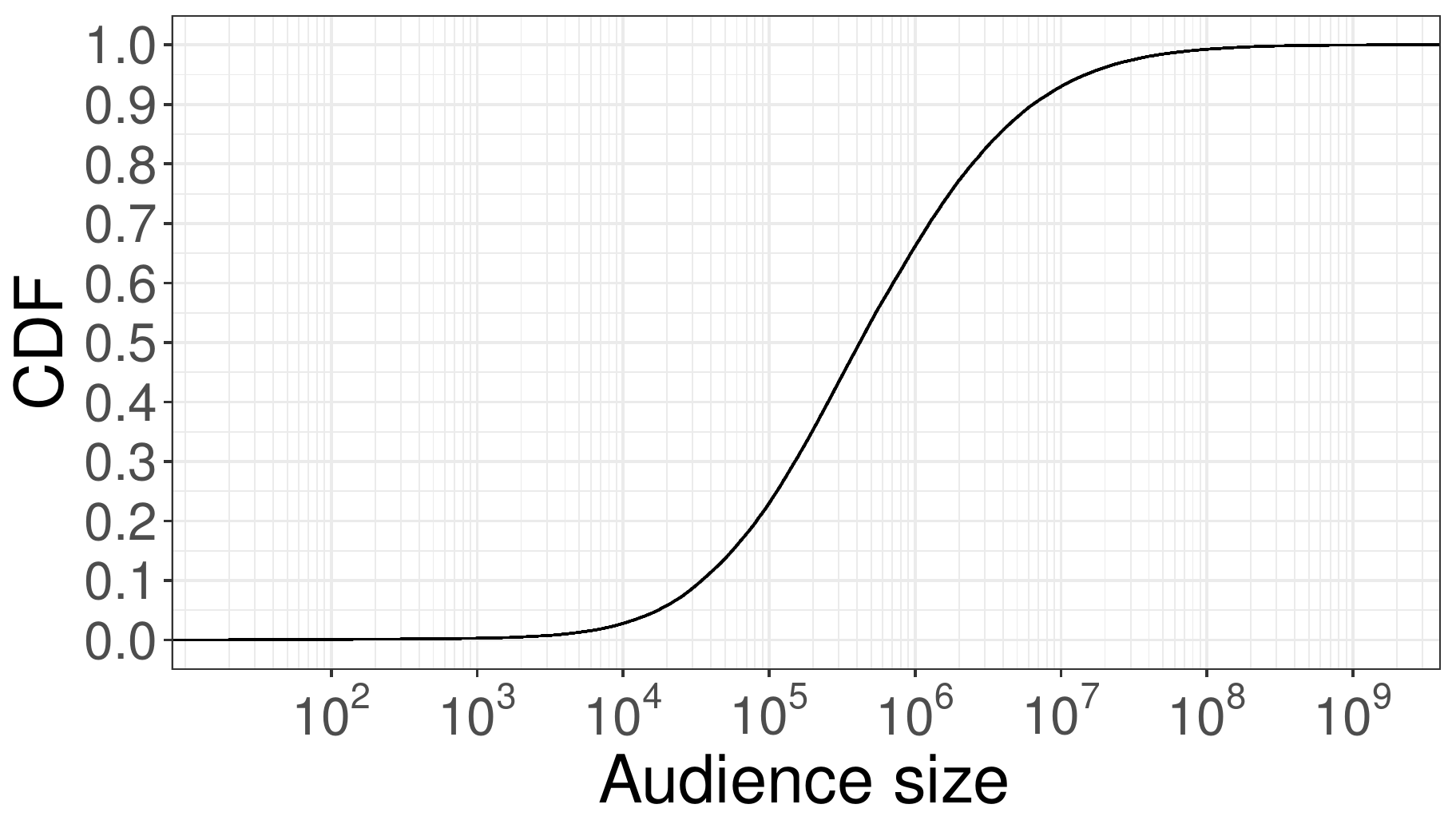}\hfill
	%\vspace{-0.2cm}
	\caption{CDF showing the distribution of the audience size for the 98,982 interests assigned to the 2,390 users of our dataset.}
	\label{fig:cdf_interests}
  \hfill
\end{figure}

To understand the popularity distribution of these interests, we extracted the audience size reported by the FB Ads Manager API for each of them. Figure \ref{fig:cdf_interests} depicts the CDF of the audience size distribution for the 99k unique interests in our dataset. The results show a large variability in the popularity of the interests. In particular the 25th, 50th and 75th percentiles of the distribution are 113,193; 418,530; and 1,719,925; respectively.

We are aware our dataset may not be a statistically \GramR{representative sample of Facebook's entire interest ecosystem}; however, it includes a very large number of interests that cover a very broad popularity range, which is what we need for the purpose of our work.  As the results in the paper validate, the collected dataset is appropriate to (i) quantify how many interests make a user unique in FB, and (ii) demonstrate that nanotargeting can be systematically implemented on FB.

%The least popular interest had an audience size of 20 with a median of 420k. %If we take the audience size of the least popular and median interest for each of the users in the dataset, in median the audience size of the least popular interest is 5.7k, and 26M the median interest.
\section{Analysis of FB User Uniqueness}
\label{sec:uniqueness}

% \begin{figure}[t]
% \centering
% 	\includegraphics[width=1\columnwidth]{./figures/synthetic_final.eps}\hfill
% 	\caption{Audience size distribution for synthetic interests ranging from 1 to 25.}
% 	\label{fig:syn_users_juan}
%   \hfill
% \end{figure} 

%The main goal of this paper is to prove the feasibility of nanotarget a user on FB based on their interests. By nanotargeting, we refer to the ability to launch an ad campaign that exclusively reaches the targeted user. To accomplish that research challenge we first need to answer the question: "how many interests unequivocally identify a user on FB?" since an advertiser/attacker will succeed in nanotargeting a user if they manage to select a set of interests leading to a FB audience size equal to 1, i.e., these interests make the targeted user unique on FB. Therefore, 
We devote this section to analyze users' uniqueness on FB according to their interests. The outcome of this section will serve two different purposes: (i) we will answer the first research question addressed in this paper: \textit{how many interests are required to uniquely identify a user on FB?}, (ii) the answer to this question will be used as a reference for the number of interests we should consider in our nanotargeting experiment (see Section \ref{sec:proof_of_concept}).

\subsection{Methodology}

We define the variable $N_P$ as the number of interests that uniquely identify a user with a probability $P$ on FB. For instance, if with 9 (18) interests a user can be uniquely identified on FB with a probability 0.3 (0.8), then $N_{0.3}$ = 9 ($N_{0.8}$=18). 

Our goal is to propose a model that defines $N_P$ for any value of $P$. We use as data source the 99k unique interests assigned to the 2,390 users of \CamR{the FDVT} browser extension.

Let us consider a user in our dataset $u_i$ (i $\in$ [1, 2390]) and a given number of interests $N$ ($N$ $\in$ [1,25]).\footnote{This range is due to the limitation imposed by the FB API that allows retrieving audience sizes for a combination of at most 25 interests.} For each pair ($u_i$, $N$), we select a set of $N$ interests from the list of interests FB assigned to $u_i$ and collect the FB audience size associated with that combination of interests leveraging the FB Ads Manager API.  
After doing this for all combinations of $u_i$ and $N$, we obtain 25 vectors, one per each value of $N$, including 2,390 audience size samples.\footnote{We note some of the vectors include fewer than 2,390 samples because in our dataset we find users that are assigned less than 25 interests. The shorter vector is the one associated with $N=25$ that includes 2,286 samples.} For instance, in the case of $N=5$, we create a vector with 2,390 audience size values retrieved from 2,390 different combinations of 5 interests (one per user in our dataset). 

Using these vectors we can produce a distribution of the audience size for each value of $N$ and compute the different quantiles of the distribution. Based on this, we define $AS(Q,N)$ as the audience size for quantile $Q$ and the number of interests $N$. For instance, an $AS(50,5)$ = 500 means that with a probability of 50\% the size of an audience defined with 5 interests is $\leq500$. Note that, since the minimum audience size reported by FB is 20, $AS(Q,N) \geq$ 20 by definition. 

Next, we create a vector $V_{AS}(Q)$ including the values of $AS(Q,N)$ for a fixed value of $Q$ and all values of $N$ (from 1 to 25). $V_{AS}(Q)$ is defined as:

\[V_{AS}(Q) = [AS(Q,1),\,AS(Q,2),\,...,\,AS(Q,24),\,AS(Q,25)]\]

\begin{figure}[t]
\centering
	\includegraphics[width=1\columnwidth]{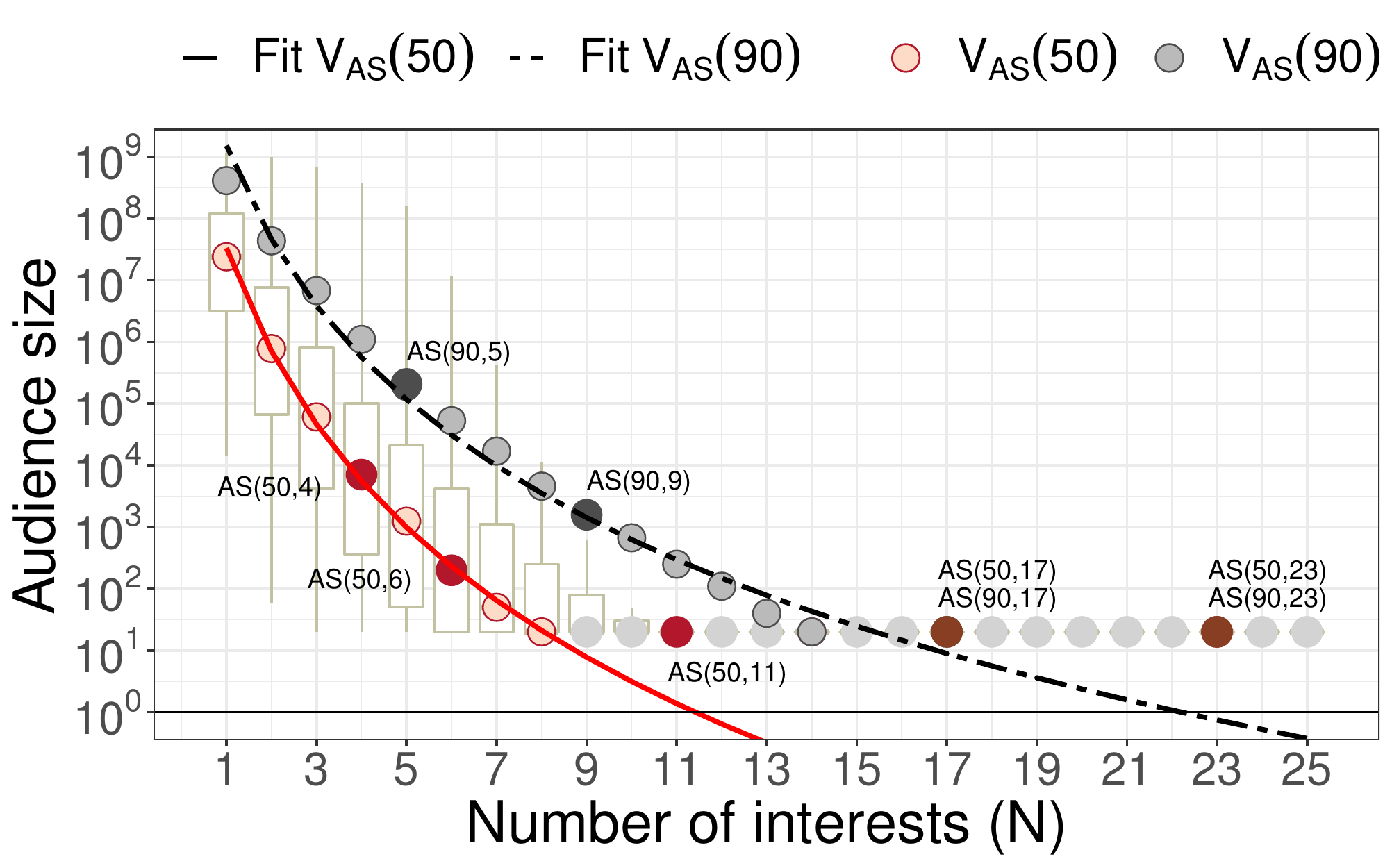}\hfill
	%\vspace{-0.2cm}
	\caption{This figure represents an illustration of our model to compute $N_p$. First, it showcases an example of the variables $V_{AS}(Q)$ and $AS(Q,N)$ for $Q$ = 50 (red dots) and $Q$ = 90 (black dots). $AS(Q,N)$ collides for both cases from N=14 when the audience size value becomes 20 (the limit imposed by FB). Second, the figure illustrates the logarithmic fitting model used to estimate the value of $N_P$ for $V_{AS}(50)$ (red line) and $V_{AS}(90)$ (black dashed line) as the cutpoint of the lines with the value y=1 (audience size equal to 1).}
	\label{fig:vas_qn}
  \hfill
\end{figure}

Since $AS(Q,N)$ $\geq$ $AS(Q,N+1)$, $V_{AS}(Q)$ presents a decreasing trend. Figure \ref{fig:vas_qn} shows examples of $V_{AS}(Q)$ for $Q$ = 50 and $Q$ = 90, where the y-axis represents the audience size and the x-axis represents the number of interests $N$. 

In the described model, \textit{$N_P$ is defined as the cutpoint where $V_{AS}(Q)$ intercepts an audience size equal to 1}. Unfortunately, as we can observe, $V_{AS}(Q)$ has an asymptote at 20 since this is the minimum audience size reported by FB. 

To overcome this issue, we fit $V_{AS}(Q)$ using the following logarithmic model: 

\[ log(V_{AS}(Q)) \sim - A log(N+1) + B \]

Based on this fit we calculate the cutpoint of the number of interests $N$ at which the regression line intercepts an audience size of 1, %This means, for a user to be unique the associated audience size should be equal to 1,
i.e.,  $V_{AS}(Q)$ = 1. Since we are using a logarithmic model the cutpoint actually happens where $log(V_{AS}(Q))$ = 0. Therfore, $N_p$ is defined as follows:

\[ N_p \geq 10^{B/A} - 1 \]

To assess the uncertainty of this estimate, we repeat the data aggregation and model fit in 10,000 bootstrap samples, calculating this way the 95\% Confidence Interval (CI) of the cutpoint for each value of $N$. Note that, we did not truncate the data for audiences of size 20, and we included the first $AS(Q,N)$ = 20 in our estimation. By doing so, our estimate of the cutpoint is conservative but robust to the minimum size of 20 and our method can still be applied for the current higher limit of 1,000 users.

Figure \ref{fig:vas_qn} shows the result of the fitting process of $V_{AS}(Q)$ for $Q$ = 50 (red dashed line) and $Q$ = 90 (black dashed line). %in an illustrative example that uses data extracted from our FDVT dataset. 

%\vspace{0.2cm}
Using as reference the outcome of the presented model, we have implemented an experiment running real FB campaigns to nanotarget three of the authors of this paper to validate: (i) whether it is feasible to implement a systematic nanotargeting attack on FB based on users' interests; (ii) if the values of $N_P$ derived from our methodology can be used as a good reference of the success probability of a potential nanotargeting attack. We describe the experiment in detail and present the obtained results in Section \ref{sec:proof_of_concept}.

\subsection{Interests Selection}

The value of $N_P$, i.e., the number of interests that make a user unique on FB with probability $P$, very much depends on the strategy used to select the interests. 

The popularity, i.e., audience size, of FB interests is very diverse and so is the popularity of the interests of an individual user. Our dataset reveals that, in general, across the hundreds of interests typically assigned to an individual user on FB, some are very popular (with audience sizes in the order of tens or hundreds of millions of users) whereas others are unpopular (with audience sizes in the order of tens or a few hundreds of users). 

The best alternative to succeed in nanotargeting a user consists in running an ad campaign selecting the least popular interests of that user as target audience, which is expected to lead to small values of $N_P$ (even for high values of $P$ like 0.9 or 0.95). However, implementing this attack would require having full knowledge of the list of interests of the targeted user, which in practice is very unlikely. Instead, it is more likely that an attacker knows a subset of the interests of the targeted user, but not all.

Having in mind that FB imposes a limitation of 25 interests in the definition of target audiences, in this paper, we apply two different approaches for the interests selection based on the previous discussion:

%The probability that a user becomes unique on FB will very much depend on the popularity (i.e., audience size) of the interests used to define the audience. For instance, the audience size associated with the interests tuple \textit{\{Sports, Coffee, Music\}} (\emph{Potential Reach}: 580,000,000 users worldwide) is much larger than the audience size of the tuple \textit{\{Roof garden, Chandler Bing, Clash Royale\}} (\emph{Potential Reach}: fewer than 1000 users worldwide), and all the six interests could be part of the interests' set of a user on FB. Therefore, the user will become unique faster (i.e., using fewer interests) if the interest set used includes the second tuple instead of the first one.

%Our dataset reveals that FB users are usually assigned hundreds or even thousands of interests. However, to compute $N_X$ we can only use combinations of up to 25 interests due to the limitation imposed by FB in the definition of audiences in advertising campaigns. 

%Therefore, the approach we use to select the interests applied to our methodology will heavily impact the actual result of $N_X$. We have used two different approaches to organize users' interests and have created two different datasets for our methodology:

\begin{itemize}
    \item {\textit{Least popular interests selection (LP):}}  We retrieve the audience size of all the interests assigned to a user and select the 25 least popular ones. We start retrieving the audience size for the least popular interest and keep adding the following least popular interests sequentially one by one to retrieve all the associated audience sizes until we complete the longest combination of 25 interests. From now on in the paper, we will use the variable $N(LP)_P$ when $N_P$ is computed selecting the least popular interests from the users.
    
    \item {\textit{Random interests selection (R):}} We select 25 interests at random from the interests assigned to a user. We start retrieving the audience size for a random interest and keep adding interests sequentially one by one to retrieve all the associated audience sizes until we complete the longest combination of 25 interests. From now on in the paper, we will use the variable $N(R)_P$ when $N_P$ is computed selecting the random interests from the users.
    
    %In this dataset we select 25 interests at random from the interests assigned to a user. We start retrieving the audience size for a random interest and keep adding interests at random until we complete the maximum value of 25 interests.
\end{itemize}

%As a result, our methodology will provide different values of $N_X$ depending on whether we apply the least popular or the random selection option. This has some interesting theoretical and practical implications. 

%The obtained $N(LP)_Q$ and $N(R)_Q$ have interesting theoretical and practical implications. On the one hand, 

The value of $N(LP)_P$ has an important theoretical relevance since it establishes a theoretical lower bound in terms of privacy based on the number of interests that make a user unique among 1.5B FB users (roughly 1/5 of the worldwide population) considered in our uniqueness analysis. Therefore $N(LP)_P$ is to the best of our knowledge the closest computation made so far concerning the number of non-PII items that make an individual unique within the whole of humanity.

However, as discussed above, $N(LP)_P$ only serves as a reference for nanotargeting purposes in those cases where the attacker knows the full list of interests of a user, which is expected not to be a common situation. Therefore, we will use $N(R)_P$ as a reference for our nanotargeting experiment introduced in Section \ref{sec:proof_of_concept}.

%On the other hand, \emph{Least Popular Interests} have less practical significance for the case of nanotargeting because it is very unlikely a third-party/attacker knows the least popular interests of a user, and in most cases, the third-party/attacker will infer a group of interests that are not necessarily the least popular. Therefore, the results obtained from the \emph{Random Interests} selection will be the ones used as a reference for our nanotargeting proof-of-concept experiments.

\subsection{Results}
\label{sec:results}

In this section, we apply the developed model to compute $N_P$, the number of interests that make a user unique on FB with a probability $P$. In particular, to perform a comprehensive discussion we consider $P$ = {0.5, 0.8, 0.9 and 0.95} and the two defined interests selection approaches: \emph{Least Popular ($N(LP)_P$)} and \emph{Random ($N(R)_P$)} interests.

We have also computed $N(LP)_P$ and $N(R)_P$ across different demographic groups based on gender, age, and location (country) to explore differences in the number of interests that make a user unique across these groups. The results of the demographic analysis are described in Appendix \ref{ap:demographic_analysis} since they are a side contribution to the core of this work.

%the following values of Q $N_{0.5}$, $N_{0.8}$, $N_{0.9}$ and $N_{0.95}$ for both (i) the \emph{Least Popular Interests} dataset, and (ii) the \emph{Random Interests} dataset.

\subsubsection{$N(LP)_P$: Least Popular Interests Selection}

\begin{figure}[!t]
\centering
%\noindent\begin{minipage}[t]{\columnwidth}
	\includegraphics[width=1\columnwidth]{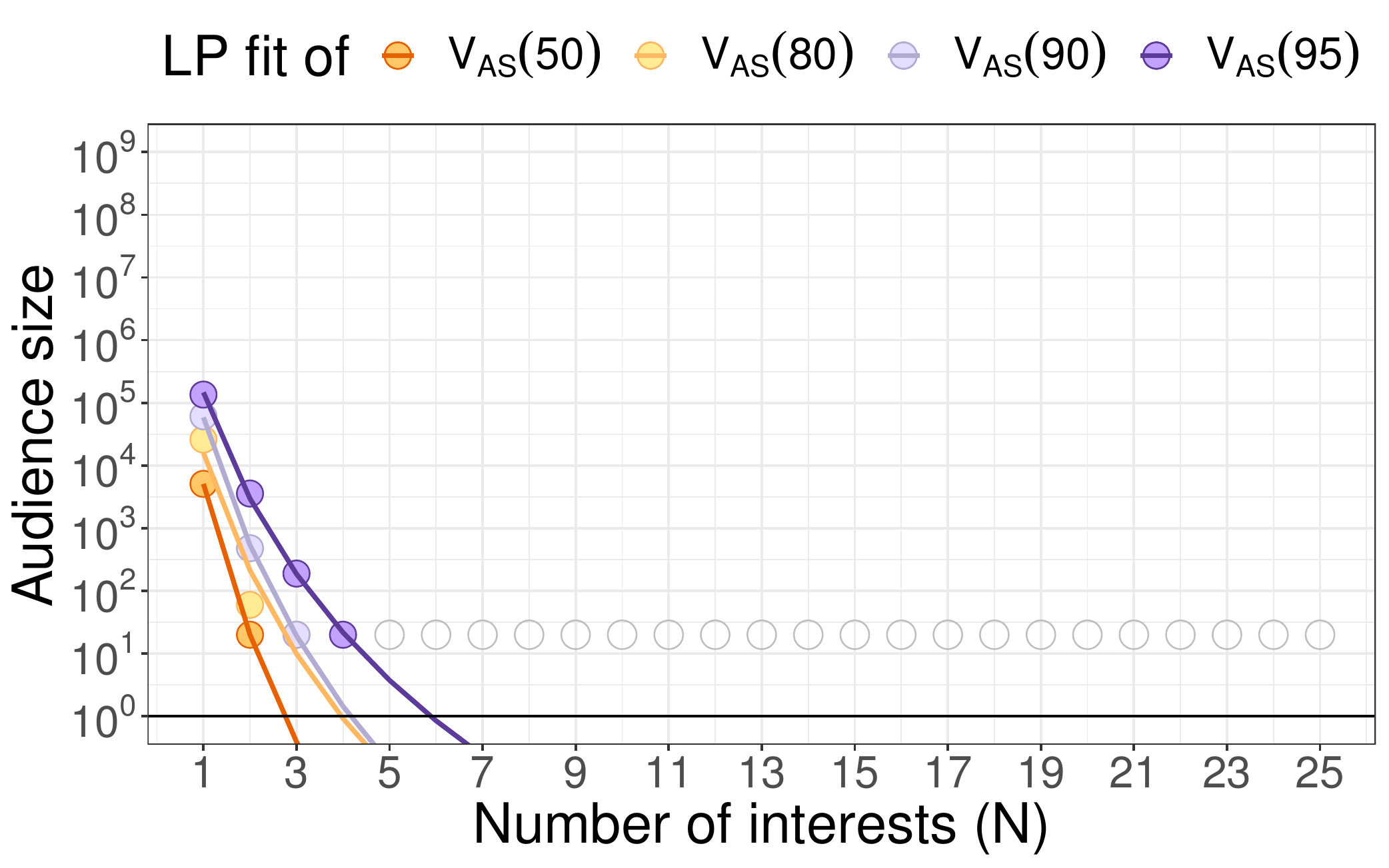}\hfill
	%\vspace{-0.2cm}
	\caption{The figure illustrates the results from our model to compute the number of interests that make a user unique on FB using their least popular interests. In particular, the figure shows the results for $N(LP)_{0.5}$, $N(LP)_{0.8}$, $N(LP)_{0.9}$ and $N(LP)_{0.95}$ applying our fitting model to the vectors $V_{AS}(50)$, $V_{AS}(80)$, $V_{AS}(90)$ and $V_{AS}(95)$. }
	\label{fig:all_least_popular}
\end{figure}

Figure \ref{fig:all_least_popular} displays $V_{AS}(Q)$ for the selection of the least popular interests of users in our dataset and $Q$ = {50, 80, 90 and 95} along with their correspondent linear fitting curve. 

Moreover, Table \ref{tab:descriptive} presents the estimated value of $N(LP)_{0.5}$, $N(LP)_{0.8}$, $N(LP)_{0.9}$ and $N(LP)_{0.95}$ along with the 95\% confidence interval and the R-squared ($R^2$) value. Both quality metrics suggest that the fitting model is very accurate. %The table also provides the results for the different considered demographic groups based on age, gender, and country. 

As discussed above, the obtained $N(LP)_P$ values offer a lower bound about the number of non-PII items that make a user unique among  1.5B FB users, roughly 1/5 of the worldwide population. $N(LP)_{0.95}$ = 5.89, indicates that a user can be uniquely identified on FB based on its 6 least popular interests with a 95\% probability. Similarly, $N(LP)_{0.9}$ = 4.16 and $N(LP)_{0.5} =$ 2.74 show that with roughly the 4 and 3 least popular interests an individual can be uniquely identified among 1.5B users with a probability of 90\% and 50\%, respectively. 

The results indicate that the number of non-PII data items that make a user unique in a worldwide population-scale dataset is really small (4 with a 90\% probability). In other words, the privacy of a user is only bounded by a handful of non-PII items. % Also, if we compare this result with previous studies conducted on datasets including around one million users\cite{de2013unique}\cite{de2015unique}, we conclude that the items of information that make a user unique in a group of billions of users are the same (4 FB interests) compared to a group with millions of users (4 credit card transactions or mobile phone calls).\ac{ESTO DIFIERE DE LO QUE CONTAMOS EN LA INTRO. EN PRINCIPIO NO ES JUSTO COMPARAR LEAST POPULAR INTERESTS CON LAS LLAMADAS PORQUE EN EL PAPER DE SCIENCE HASTA DONDE SABEMOS NO USAN LAS LEAST POPULAR. HAY QUE ELEGIR LA HISTORIA QUE QUEREMOS CONTAR Y SER CONSISTENTES.} %This opens avenues for further research in other disciplines such as ethics, sociology, or economy. 

Finally, in the context of nanotargeting, our result suggests that an attacker having full access to the list of interests of a user can nanotarget them with a 90\% probability by running an ad campaign with just 4 interests. The success probability increases to 95\% if the attacker uses the 6 least popular interests.

\begin{table*}[!t]
\resizebox{\textwidth}{!}{%
\begin{tabular}{l||l|l|l||l|l|l||l|l|l||l|l|l}

$\mathbf{N_P}$ & \textbf{P=0.5} & \textbf{95\% CI} & $\mathbf{R^2}$ & \textbf{P=0.8} & \textbf{95\% CI} & $\mathbf{R^2}$ & \textbf{P=0.9} & \textbf{95\% CI} & $\mathbf{R^2}$ & \textbf{P=0.95} & \textbf{95\% CI} & $\mathbf{R^2}$ \\ \hline
    $N(LP)_P$ & 2.74 & (2.72,2.75) & 1.00 & 3.96 & (3.91,4.02) & 0.92 & 4.16 & (4.09,4.37) & 1.00 & 5.89 & (5.62,6.15) & 1.00 \\ 
    $N(R)_P$  & 11.41 & (11.21,11.6) & 1.00 & 17.31 & (16.98,17.6) & 0.99 & 22.21 & (21.73,22.69) & 0.99 & 26.98 & (26.34,27.68) & 0.98 \\ 
   \hline
\end{tabular}%
}
\caption{Number of interests needed to make a user unique on FB with probability 0.5, 0.8, 0.9 and 0.95 ($N_{0.5}$, $N_{0.8}$, $N_{0.9}$ and $N_{0.95}$). The first row reveals the results for the case in which we select the least popular users' interests (i.e., $N(LP)_P$). The second row exposes the results for a random selection of users' interests (i.e., $N(R)_P$). We provide the results along with the 95\% Confidence Interval (CI) and the R-squared (R\textsuperscript{2}) associated with the fitting model used to obtain $N(LP)_P$ and $N(R)_P$.}
\label{tab:descriptive}
\end{table*}

\subsubsection{$N(R)_P$: Random Interests Selection}

\begin{figure}[!t]
\centering
	\includegraphics[width=1\columnwidth]{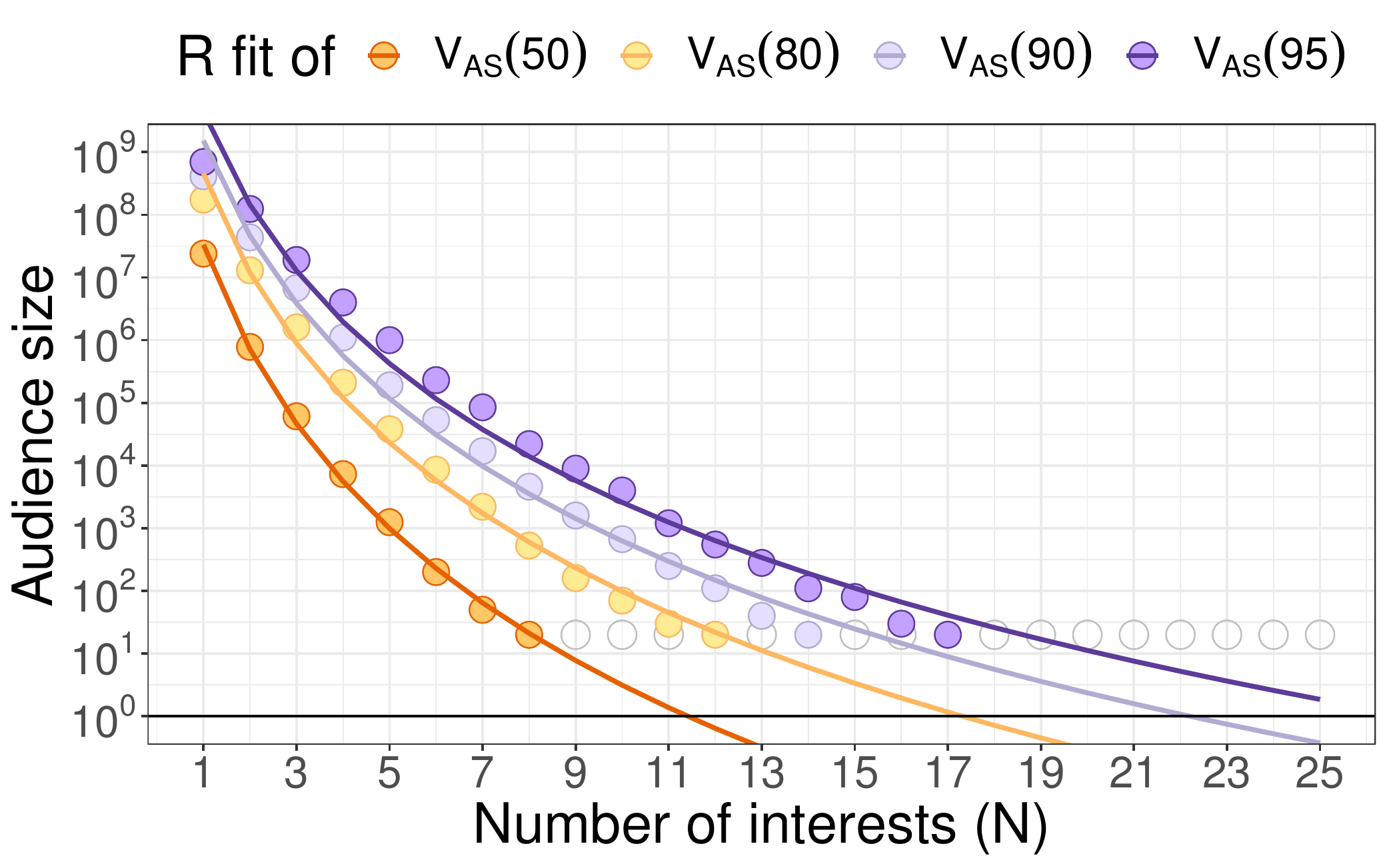}\hfill
	%\vspace{-0.2cm}
	\caption{The figure illustrates the results from our model to compute the number of interests that make a user unique on FB combining interests at random. In particular, the figure shows the results for $N(R)_{0.5}$, $N(R)_{0.8}$, $N(R)_{0.9}$ and $N(R)_{0.95}$ applying our fitting model to the vectors $V_{AS}(50)$, $V_{AS}(80)$, $V_{AS}(90)$ and $V_{AS}(95)$. }
	\label{fig:all_random}
\end{figure}

In this subsection, we present an analysis for $N(R)_P$ similar to the one presented above for $N(LP)_P$. Figure \ref{fig:all_random} shows $V_{AS}(Q)$ based on the random selection of users' interests and $Q$ = {50, 80, 90 and 95} along with their correspondent linear fitting curve. 

In addition, Table \ref{tab:descriptive} presents the obtained estimations for $N(R)_{0.5}$, $N(R)_{0.8}$, $N(R)_{0.9}$, and $N(R)_{0.95}$ along with their associated confidence intervals and R-squared ($R^2$) value. The confidence intervals and R-squared values indicate again a good accuracy of the proposed model.  %Values for the different considered demographic groups are also included in the Table. 

The obtained results reveal that 12, 18, 22, and 27 random interests make a user unique on FB with a probability of 50\%, 80\%, 90\%, and 95\%, respectively. These findings have two main practical implications: 1) Given that FB typically assigns hundreds of interests to users, it is likely that an attacker can infer a few tens of those interests which would enable him to nanotarget the victim; 2) Performing an attack with 95\% success probability is impossible in practice since it requires to target an audience combining 27 interests when FB imposes a maximum of 25 interests for a targeted audience. 
Note that we use these $N(R)_P$ values as a reference to run the nanotargeting experiment in Section \ref{sec:proof_of_concept}.

\hfill \break
\noindent\emph{\underline{{Summary of uniqueness analysis results.}}}
\hfill \break
\emph{Our results reveal that: $(i)$ the 4 rarest interests of a user make them unique within a user base in the same order of magnitude as the worldwide population. This indicates that the uniqueness of an individual is defined by a small number of non-PII items, thus users privacy is very compromised in the current hyper-connected society; $(ii)$ In comparison with previous studies, our analysis reveals that the number of non-PII items that make unique a user in a dataset with millions of users (4 credit card purchases or 4 mobile phone calls) or billions of users (4 rarest interests or 22 random interests) is in the same order of magnitude. This means that belonging to larger-scale human groups does not seem to contribute to significantly improve the privacy boundaries for individuals.} %$(iii)$ Finally, our demographic analysis \jg{(Appendix \ref{ap:demographic_analysis})} reveals that women, adolescents, and users from Argentina (compared to France, Spain, and Mexico) are better protected from nanotargeting attacks based on random interests selection.}

\section{Nanotargeting Experiment}
\label{sec:nanotargeting_experiment}
%\begin{table}[ht]
%\resizebox{.45\textwidth}{!}{
%\begin{tabular}{rrrrrrrrr}
%  \hline
% \#interest & 50th pctl & 80th pctl & 85th pctl & 90th pctl \\ 
%  \hline
%  5 & 1455.06 & 17540.87 & 35432.07 & 91703.69 \\ 
%  7 & 132.11 & 1600.14 & 3207.56 & 9542.92 \\ 
%  9 & 20.55 & 249.78 & 497.72 & 1649.82 \\ 
%  12 & 2.30 & 28.13 & 55.66 & 209.51 \\ 
%  18 & 0.10 & 1.20 & 2.34 & 10.59 \\ 
%  20 & 0.04 & 0.52 & 1.02 & 4.82 \\ 
%  22 & 0.02 & 0.24 & 0.47 & 2.36 \\ 
%   \hline
%\end{tabular}
%}
%\caption{Number of users reached estimated by our model.}
%\label{tab:model_estimations}
%\end{table}

In this section, we present an experiment that builds on top of the results obtained from our analysis in the previous section. Our goal is to provide evidence that the FB advertising platform can be systematically exploited to implement nanotargeting campaigns with non-PII data nowadays.

%We have implemented a proof-of-concept experiment that aims at nanotargeting 3 of the authors of this paper on FB based on the interests FB has assigned them. The goal of this proof-of-concept exercise is twofold: (i) proving that nanotargeting real-users on Facebook is feasible, (ii) evaluating whether the model we have introduced in Section \ref{sec:methodology_dataset} is valid to (roughly) predict the success probability of a nanotargeting attack. %The total cost of the experiment was 309€.

We remind that our definition of nanotargeting requires that the ad is exclusively delivered to the targeted user. Therefore, when we indicate that a nanotargeting campaign has failed it does not imply the campaign did not reach the targeted user. It means that more than one user has been reached, which may or may not include the targeted user. 

\subsection{Description of the Experiment}

The experiment consisted in creating tailored ad campaigns on FB to reach three authors of this paper using random sets of interests obtained from the complete list of interests FB had assigned them. As discussed above, we decided to focus our experiment on the use of random interests since in practice it is much more likely that an attacker knows a random list of interests from a user than their least popular interests.

\paragraph{Interests selection.}
Based on the results presented in Section \ref{sec:uniqueness}, we have run 7 campaigns per user configured with {5, 7, 9, 12, 18, 20 and 22} randomly selected interests. Since we are targeting 3 independent users, we run 21 FB ad campaigns in total in our experiment. 

We divide the experiments into two groups based on the expected success likelihood. The first group includes those experiments using {12, 18, 20, and 22} interests. We refer to it as \emph{Success Group} because our results in Section \ref{sec:uniqueness} indicate that the probability of succeeding in a nanotargeted campaign for the considered number of interests in this group ranges between 50\% and 90\%. Hence, we expect that many of these campaigns effectively nanotarget the correspondent user.

The second group configures campaigns using {5, 7, and 9} interests and we refer to it as \emph{Failure Group} since the results from our model manifest a success probability of 2.5\%, 15\% and 30\% for 5, 7, and 9 interests, respectively. Based on this, we expect most of these nanotargeting campaigns fail. 

To conduct the experiments, we select a random set of 22 interests from each user that are directly used in the campaigns configured with 22 interests. To create the campaign using 20 interests, we remove 2 from the initial set. Similarly, to create the campaign with 18 interests, we remove 2 from the set used in the 20-interest campaign. Following the same approach, we remove 6 interests from the 18-interest campaign and use the remaining ones to define a 12-interest campaign. We keep on the same process to define the remaining campaigns.

\paragraph{Geographical range.}

The geographical range of the ad campaigns is defined as ``worldwide''. This makes that our campaigns do not filter users based on their location and thus they can potentially reach any FB user. Note, that FB reported 2.8B monthly active users in the last quarter of 2020 \cite{fb_q4_2020} when we run our experiment.

\paragraph{Ad assets.}

\begin{figure}[!t]
\centering
	\includegraphics[width=\columnwidth]{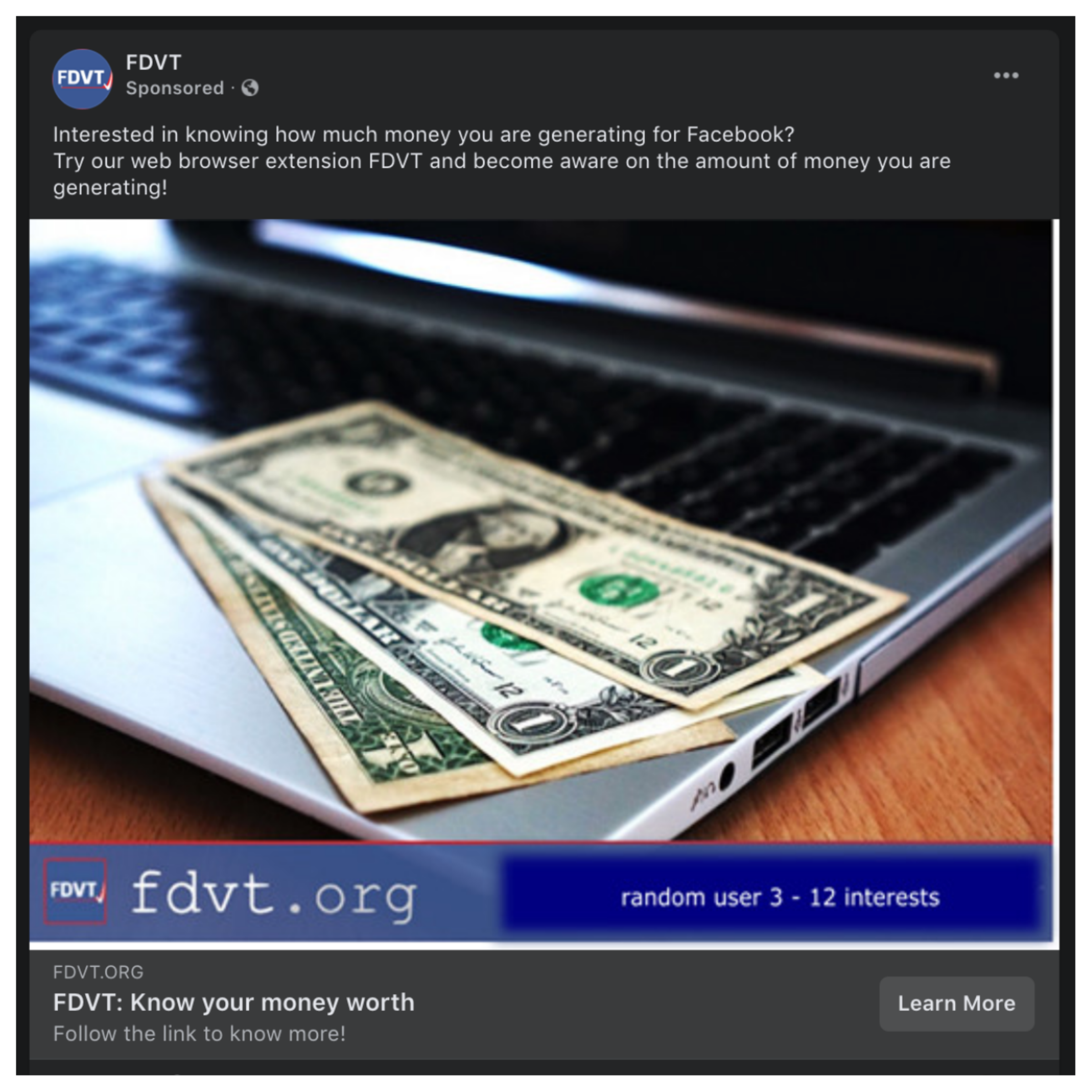}\hfill
	\caption{Ad image used in the campaign targeting User 3 with 12 interests. Every ad included a text identifying the campaign at the bottom right corner.} %For the double-blind review process, we removed the name instances of our browser add-on from the figure.
	%\caption{Ad image used in the campaign targeting User 3 with 12 interests. Every ad included a text identifying the campaign at the bottom right corner. For the double-blind review process, we have removed the name links and identifiable information of our web browser extension from the figure wherever it appeared, and hover them with white boxes.}
	\label{fig:ad_ap1}
  \hfill
\end{figure}

We created one specific ad creativity for each one of the 21 configured campaigns. Each ad creativity identifies both the user being targeted (User 1, User 2, or User 3) and the number of interests used in its associated campaign (5, 7, 9, 12, 18, 20, 22).  For instance, Figure \ref{fig:ad_ap1} presents the ad received by User 3 in his FB newsfeed associated with the nanotargeting campaign configured with 12 random interests. Moreover, each ad creativity is linked to a different landing page hosted on our web server.

%Each of the 21 ad campaigns used a different ad c that identified both the user being targeted, either User 1, User 2, or User3, and the number of interests used in that campaign. All ads included in the bottom right corner of the creativity a text indicating the user being targeted and the number of interests used in that particular campaign. Figure \ref{fig:ad_u2} illustrates the case of the ad targeting user 2 with 22 interests.

\paragraph{Timing and duration.}

Every campaign ran for a total time of 33 hours divided into 4-time windows. In particular, the campaigns in the \emph{Success Group} (using 12, 18, 20, and 22 interests) ran in parallel on Thu. Oct 29, 2020, from 19h to 21h CET, Fri. Oct 30 from 9h to 21h CET, Mon. Nov 2 from 9h to 21 CET and Tue. Nov 3 from 9h to 16h CET. We stopped all the \emph{Success Group} campaigns at the same time once the three users had received at least once the targeted ad from every campaign. The campaigns in the \emph{Failure Group} (using 5, 7, and 9 interests) ran in parallel exactly at the same hours and days as the \emph{Success Group} campaigns in the week after. 

We used the same duration, and same weekdays and hours in every ad campaign, to guarantee that all of them had the same amount of time to deliver ad impressions and to avoid potential biases in our results due to special conditions affecting concrete weekdays or hours within a day.

\paragraph{Budget.}

We allocated an initial daily budget of 70€ to each ad campaign for a period of one week. FB distributes the budget over the provided time window, so our expected expenditure was roughly 10€/day per campaign. Since the actual duration of our campaigns was lower than the provided time window of one week, none of the campaigns consumed the 70€ assigned to them. The overall cost of our experiment was 305.36€.

\paragraph{Nanotargeting success validation.}

We used three elements to validate the success (or failure) of a nanotargeted ad campaign from our experiment: 

\begin{enumerate}

\item Facebook offers advertisers a dashboard where they can monitor the progress of their ad campaigns. This dashboard reports for each campaign (among other things):  the number of delivered impressions, the number of unique users reached, the number of clicks on the ad, and the budget spent in the ad campaign.

\item We keep a record of each ad impression delivered to the targeted users. To this end, upon the reception of a targeted ad, the user was instructed to click on it. Since each ad creativity has a unique associated landing page, this click created a log entry in our web server recording the details of the ad campaign (targeted user and number of interests) and the timestamp.

%This click created a log entry in a server we managed. This log entry records the following information: user ID and the number of interests of the specific ad campaign and a timestamp. 

%receiving an ad from the nanotargeting campaigns had to click on it. That click created a log entry in a server we managed. The log entry registered: the user ID, the number of interests of the campaign, and a timestamp. 

\item Each user was also instructed to take a snapshot of the received ad along with the information included in the \textit{``Why am I seeing this ad?''} option that Facebook offers to users. When a user clicks on the \textit{``Why am I seeing this ad?''} option, a new window pops up displaying the parameters used by the advertiser to define the targeted audience in the ad campaign associated with the ad. In our experiment, those parameters refer to the list of interests used in the ad campaign. 
\GramR{Figures \ref{fig:ad_ap2} and \ref{fig:ad_ap3} in Appendix \ref{ap:ads_fdvt} illustrate an example of the \textit{``Why am I seeing this ad?''}} option captured by one of the authors. We have verified that for every targeted ad identified by the authors the parameters included in the \textit{``Why am I seeing this ad?''} matched exactly the configured audience associated with the received ad.  
\end{enumerate}

Combining the three described pieces of information we could easily identify whether a nanotargeting campaign had been successful or not. In particular,  we could safely conclude that a nanotargeted campaign had succeeded if the following three conditions hold: $(i)$ FB reported that only one user had been reached, $(ii)$ we had a log record in our web server generated by the user click in the ad, and $(iii)$ the nanotargeted user collected a snapshot of the ad and its associated \textit{``Why am I seeing this ad?''} option.

%from the targeted user who has received the ad (i.e., one of the targeted authors of the paper), we could safely conclude that this particular nanotargeting campaign had been successful.

\label{sec:proof_of_concept}

\subsection{Results}

Table \ref{tab:ad_campaigns} summarizes the results of the nanotargeting experiment. For each user and ad campaign (i.e., number of interests used) the table depicts the following five metrics: $(i)$ \textit{Seen}: it is a binary metric that indicates whether the user has received the ad or not; $(ii)$ \textit{Reached}: it reports the total number of unique users the campaign has reached based on the information reported in the FB campaign manager dashboard; (iii) \textit{Impressions}: it indicates the total number of impressions delivered by the campaign as reported by the FB campaign manager dashboard. Note that, the number of impressions is usually larger than the number of users reached because the ad can be delivered multiple times to the same user; (iv) \textit{Time to the First Impression (TFI)}: it shows the elapsed time since the campaign was launched until the first impression of the ad was received by the targeted user. To compute this metric we only consider the periods when the campaign was active; (v) \textit{Cost}: it reports the amount FB billed us in euros.  
Finally, the table highlights in bold the successful nanotargeting campaigns, i.e., the campaigns that exclusively reached the targeted user. 

In the rest of the section, we discuss the most relevant aspects of the obtained results. 

\begin{table}[t]
\resizebox{\hsize}{!}{%
\begin{tabular}{lcrrrrr}
 & \multicolumn{5}{c}{User 1} \\ \cline{2-7} 
 & Seen & \multicolumn{1}{c}{Reached} & \multicolumn{1}{c}{Impressions} & \multicolumn{1}{c}{TFI} & \multicolumn{1}{c}{Cost} & \multicolumn{1}{c}{\CamR{Clicks}} \\ \hline
\multicolumn{1}{l|}{5 interests} & No & 9,824 & 42,273 & - & €28.58 & \CamR{40 (38)} \\
\multicolumn{1}{l|}{7 interests} & No & 2,992 & 14,774 & - & €29.47 & \CamR{14 (13)}\\
\multicolumn{1}{l|}{{9 interests}} & {Yes} & {743} & {4,883} & {2h 11'} & {€28,74} & \CamR{17(14)} \\
\multicolumn{1}{l|}{{12 interests}} & {Yes} & {152} & {1,110} & {9h 8'} & {€19.28} & \CamR{9 (4)}\\
\multicolumn{1}{l|}{\textbf{18 interests}} & \textbf{Yes} & \textbf{1} & \textbf{1} & \textbf{3h 31'} &\textbf{€0.01} & \CamR{\textbf{1 (1)}}\\
\multicolumn{1}{l|}{\textbf{20 interests}} & \textbf{Yes} & \textbf{1} & \textbf{1} & \textbf{47'} & \textbf{Free} & \CamR{\textbf{1 (1)}}\\
\multicolumn{1}{l|}{\textbf{22 interests}} & \textbf{Yes} & \textbf{1} & \textbf{1} & \textbf{28h 40'} & \textbf{Free} & \CamR{\textbf{1 (1)}}\\ \hline \\
 & \multicolumn{5}{c}{User 2} \\ \cline{2-7} 
 & Seen & \multicolumn{1}{c}{Reached} & \multicolumn{1}{c}{Impressions} & \multicolumn{1}{c}{TFI} & \multicolumn{1}{c}{Cost} & \multicolumn{1}{c}{\CamR{Clicks}} \\ \hline
\multicolumn{1}{l|}{5 interests} & No & 89,328 & 251,379 & - & €28.97 & \CamR{94 (94)}\\
\multicolumn{1}{l|}{{7 interests}} & {Yes} & {1,843} & {10,004} & {2h 9'} & {€29.30} & \CamR{23 (22)} \\
\multicolumn{1}{l|}{{9 interests}} & {Yes} & {1,152} & {7,175} & {1h 47'} & {€29.00} & \CamR{19 (19)} \\
\multicolumn{1}{l|}{{12 interests}} & {Yes} & {201} & {970} & {4h 22'} & {€18.68} & \CamR{11 (6)} \\
\multicolumn{1}{l|}{{18 interests}} & {Yes} & {92} & {263} & {27h 57'} & {€4.15} & \CamR{6 (3)}\\
\multicolumn{1}{l|}{\textbf{20 interests}} & \textbf{Yes} & \textbf{1} & \textbf{1} & \textbf{44'} & \textbf{€0.01} & \CamR{\textbf{1 (1)}}\\
\multicolumn{1}{l|}{\textbf{22 interests}} & \textbf{Yes} & \textbf{1} & \textbf{1} & \textbf{32h 10'} & \textbf{€0.01} & \CamR{\textbf{1 (1)}}\\ \hline \\
 & \multicolumn{5}{c}{User 3} \\ \cline{2-7} 
 & Seen & \multicolumn{1}{c}{Reached} & \multicolumn{1}{c}{Impressions} & \multicolumn{1}{c}{TFI} & \multicolumn{1}{c}{Cost} & \multicolumn{1}{c}{\CamR{Clicks}} \\ \hline
\multicolumn{1}{l|}{5 interests} & No & 39,520 & 100,106 & - & €30.05 & \CamR{93 (90)} \\
\multicolumn{1}{l|}{7 interests} & No & 2,221 & 11,248 & - & €30.83 & \CamR{26 (25)}\\
\multicolumn{1}{l|}{{9 interests}} & {Yes} & {749} & {4,356} & {1h 50'} & {€28,19} & \CamR{22(15)} \\
\multicolumn{1}{l|}{\textbf{12 interests}} & \textbf{Yes} & \textbf{1} & \textbf{1} & \textbf{12h 22'} & \textbf{€0.01} & \CamR{\textbf{1 (1)}}\\
\multicolumn{1}{l|}{\textbf{18 interests}} & \textbf{Yes} & \textbf{1} & \textbf{2} & \textbf{6h 19'} & \textbf{€0.02} & \CamR{\textbf{2 (2)}}\\
\multicolumn{1}{l|}{\textbf{20 interests}} & \textbf{Yes} & \textbf{1} & \textbf{5} & \textbf{3h 32'} & \textbf{€0.06} & \CamR{\textbf{5 (3)}} \\
\multicolumn{1}{l|}{\textbf{22 interests}} & \textbf{Yes} & \textbf{1} & \textbf{1} & \textbf{48'} & \textbf{Free} & \CamR{\textbf{1 (1)}}\\ \hline
\end{tabular}%
}
\caption{Results of the nanotargeting experiment for three authors of the paper. The rows indicate the number of interests used in each of the 7 ad campaigns launched per user. The columns represents the performance: \textit{Seen} (whether the targeted user received the ad or not); \textit{Reached} (the number of users reached by the campaign); \textit{Impressions} (the total number of impressions delivered in the campaign); \textit{TFI} (time to the first impression delivered to the targeted user); \textit{Cost} (cost of the campaign); \CamR{\textit{Clicks} (the number of clicks in the campaign and number of pseudonymized unique IP addresses generating those clicks, in parenthesis)}.}
\label{tab:ad_campaigns}
\end{table}

\paragraph{Nanotargeting feasibility.}

9 out of the 21 campaigns successfully nanotargeted the correspondent user. These campaigns are all 20-interests and 22-interests campaigns,  two (out of the three) 18-interests campaigns, and one (out of the three campaigns) using  12 interests. 
There are six other campaigns (two 12-interests, the three 9-interests, and one 7-interests) that also reached the targeted user along with other few hundreds or thousands of users. Therefore, these campaigns failed to nanotarget (i.e., exclusively reach) a specific user.
Finally, there are five campaigns (the three 5-interests and two 7-interests) that did not reach the targeted user.

In a nutshell, our experiment demonstrates that an attacker can systematically nanotarget a single user on FB if they can infer a sufficient number of interests from the individual being targeted. 

\paragraph{Cost of nanotargeting.}

An important question is what is the actual cost associated with a nanotargeted campaign. Note that a very high cost may serve as a discouraging factor in practice. Unfortunately, results extracted from the FB Ad Campaign Manager and reported in Table \ref{tab:ad_campaigns} prove that nanotargeting a user is rather cheap. Indeed, the overall cost of the 9 successful nanotargeting campaigns was only 0.12€. Surprisingly, FB did not charge us anything in three of the successful nanotargeting campaigns that delivered only 1 ad impression to the targeted user.
Therefore, rather than a discouraging factor, the extremely low cost of nanotargeting may encourage attackers to leverage this practice.  

\paragraph{Time to the First Impression.}

The results expose a wide variability of the TFI, which ranges between 44m to 32h10m across the 9 successful campaigns. In particular, 3 of these campaigns show a TFI lower than an hour whereas 3 of them present a TFI higher than 10h. 

\CamR{
\paragraph{Clicks.} 

We show the number of clicks retrieved in each ad campaign as well as the number of unique IP addresses (using the pseudonymized IP address information we stored) generating those clicks. The latter represents an upper bound of the number of users that clicked on the ad. Since we are only storing the (pseudonymized) public IP address of the device generating the click, we cannot validate whether the same user receives our ad multiple times and clicked on it from different IP addresses (e.g., different devices, same mobile device connected to different access points, etc.). For instance, User 3 clicked the 5 ad impressions from the 20 interests campaign from 3 different IP addresses. As it was expected, the successful nanotargeting campaigns only received clicks from the targeted users as the number of clicks and impressions match one to one.
}

\hfill \break
\noindent\emph{\underline{Summary of nanotargeting experiment results.}}
\hfill \break
\emph{The main conclusions derived from our experiment are the following: (i) nanotargeting a user on FB is highly likely if an attacker can infer 18+ interests from the targeted user; (ii) nanotargeting is extremely cheap, and (iii) based on our experiments, 2/3 of the nanotargeted ads are expected to be delivered to the targeted user in less than 7 effective campaign hours.}

\section{FDVT Browser Extension Solution to Remove Specific Interests}
\label{sec:fdvt_solution}

We have developed a solution that displays a list of the interests Facebook has assigned to a user sorted based on their audience size, from the lowest to the highest value. This solution: (i) informs users that some of the interests in their set may be too specific, and can be used for inappropriate privacy abusive practices such as nanotargeting, (ii) allow users to easily delete any of the interests in the list by just clicking a button. Hence, the solution offers a simple and guided mechanism for users to delete the least popular interests in their list to protect their privacy.

We have implemented this solution as a new feature in the \CamR{FDVT} browser extension we used to collect the dataset used for our uniqueness analysis in Section \ref{sec:uniqueness}. To obtain the audience size of the interests assigned to the user, each time a user starts a session on FB, the \CamR{FDVT} browser extension retrieves their updated set of ad preferences (i.e., interests) and the audience size of each interest from the FB Ads Manager API. Based on this information, the browser extension computes a sorted list of the interests assigned to the user from least to most popular. The graphical interface of the \CamR{FDVT} browser extension adds a new button with the label \textit{``Risks of my FB interests''}. When a user clicks on that button, the extension displays a web page displaying the sorted list of interests. We have defined a color code to facilitate users' understanding of what interests may lead to a major privacy risk based on their associated audience size. We use the following classification: Red (high risk) for worldwide audience sizes $\leq10k$ users; Orange (medium risk) for audience sizes between 10k and 100k users; Yellow (low risk) for audience sizes between 100k and 1M users; Green (no risk) for audience sizes $\geq1M$ users. Note that, the threshold for each risk category can be easily modified if other scientific works or experts recommend using different values.

\begin{figure}[!t]
\centering
	\includegraphics[width=\hsize]{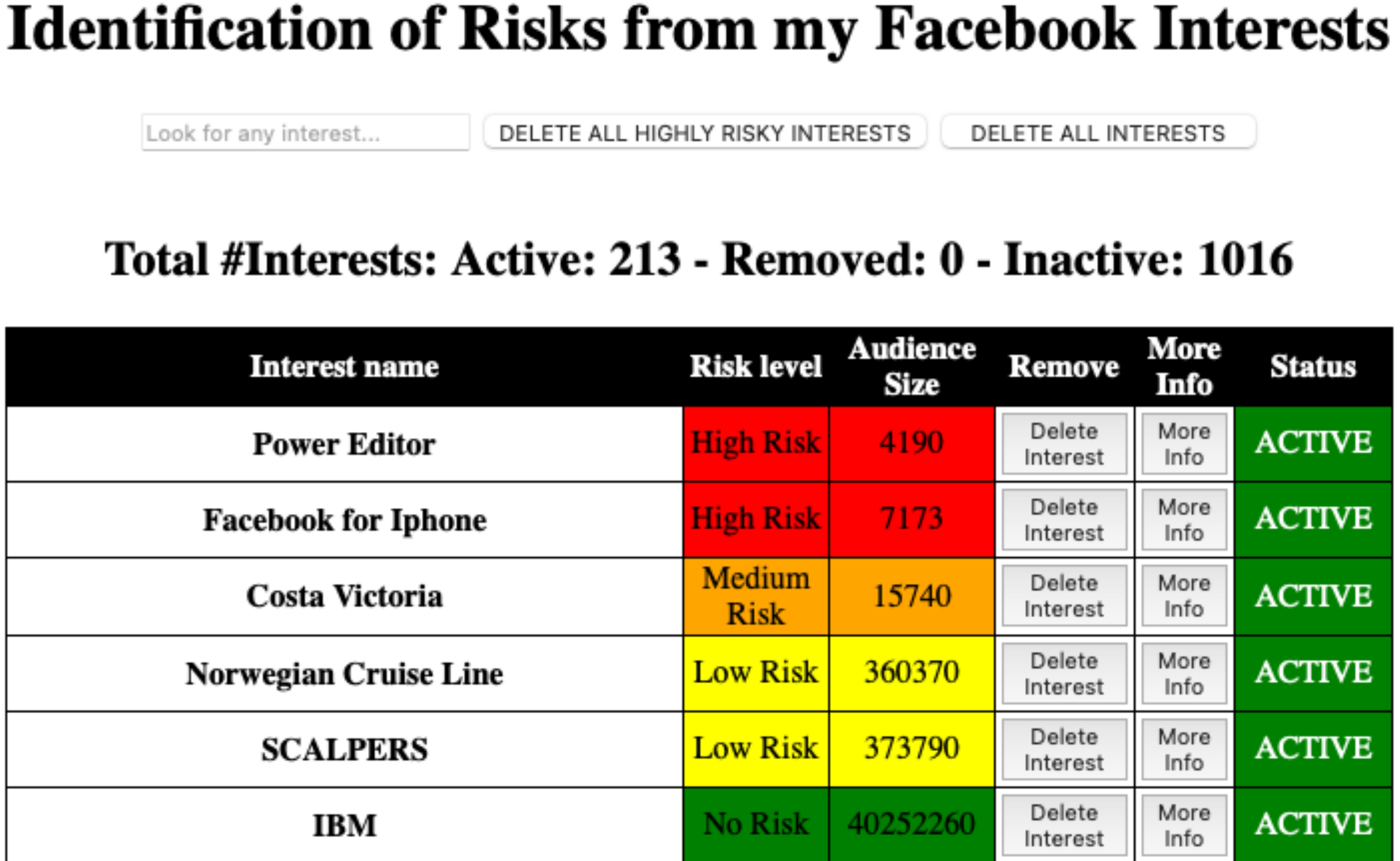}\hfill
	%\vspace{0.5cm}
	\caption{Snapshot of the interface of \CamR{the FDVT} browser extension new functionality. It informs about the potential privacy risk associated with each FB interest using a color code. It also allows the users to remove any interests with a click.}
	\label{fig:fdvt_web}
  \hfill
\end{figure}

Finally, the information displayed by this new functionality of the \CamR{FDVT} browser extension is: $(i)$ Interest name, $(ii)$ Risk level (based on the described color code), $(iii)$ Audience size, $(iv)$ Remove button, which allows deleting the associated interest from the user's profile, $(v)$ More info button, which shows historic information and the reason why that interest appears/appeared in the user's profile, and $(vi)$ Status, either active (currently in the user's ad preference set) or inactive. Figure \ref{fig:fdvt_web} depicts a snapshot of the described solution.
%Figure \ref{fig:fdvt_web} in Appendix \ref{ap:appendix_solution} shows a snapshot of the described solution.

\section{Related Work}

In this section, we present the most relevant literature for our paper in the context of users' uniqueness analyses based on non-PII data and nanotargeting experiments.

\subsection{Uniqueness Based on non-PII Items}

There is an existing body of literature that has explored the number of non-PII items required to uniquely identify a person within a large user base. Sweeney \cite{sweeney2000simple} reported that getting access to gender, ZIP code and birth date of users allowed revealing the identity of 87\% citizens within the 1990 US census data that included 248M persons. Most recently, Golle et al. \cite{Golle:2006:RUS:1179601.1179615} replicated Sweeney's analysis using the 2000 US census, which included 281M individuals. The results show a drop from 87\% to 63\% in a period of 10 years. De Montjoye et al. \cite{de2013unique} exposed that knowing the time and location associated with only four mobile calls was enough to uniquely identify 95\% of the individuals in a dataset including 1.5M users. Similarly, De Montjoye et al. \cite{de2015unique} studied 3 months of credit card records from 1.1 million people and revealed that four spatio-temporal points from credit card purchases are enough to uniquely identify 90\% of individuals in such dataset. \jg{Su et al. \cite{su} demonstrated the possibility of uniquely identify a Twitter user based on their browsing history. They performed an experiment using real users' browsing history and successfully de-anonymized 268 out of 374 real Twitter accounts (72\%)}. Finally, Narayanan et al. \cite{Netflix_Reidentification} tried to deanonymize the  Netflix Prize dataset \cite{Netflix_prize} that included more than 100M movie ratings from 480k Netflix subscribers between December 1999 and December 2005. The authors demonstrated that 8 movie ratings along with their dates (that may have a 14-day error) are enough to uniquely identify 99\% of the users in the dataset. Besides, they used a small sample of 50 (known) users from the Internet Movie Database (IMDb) \cite{IMDb} that had publicly rated movies and were able to identify two of them in the Netflix database. This demonstrates that information obtained from an online system A can be used to unveil the identity of users in an online system B.  

Our work contributes to this literature in various ways: 

(i) Our study is the first one analyzing uniqueness within a user base at the scale of the worldwide population. Indeed, the user base of our study represents around 1/5 of the world population. Previous works have used datasets from private companies including at most 1.5M or the US census with up to 281M people. However, the reidentification capacity in the 2000 US census dataset (i.e., the most recent analysis) is significantly smaller than in our study.

(ii) All previous works rely on location and/or temporal information from the users. Instead, our study considers a completely different type of non-PII item represented by the interests of users in social networks.

(iii) The pieces of information used in our study, i.e., users' interests, can be straightforwardly used to define ad campaigns in platforms like FB at very high reidentification rates (i.e., 90\%). In contrast, previous works either rely on information from private companies (call registers or credit card transactions) that is not directly actionable to target an individual or achieve a rather low reidentification rate (63\% in the US 2000 census study).

\subsection{Nanotargeting}

Researchers and practitioners have explored the possibility of implementing nanotargeting on FB. Existing literature can be classified into two groups according to the FB tools used to implement nanotargeting campaigns. On the one hand, and similarly to our approach, a couple of preliminary studies address the implementation of nanotargeted campaigns using the standard FB Ads Manager dashboard and non-PII data. On the other hand, we find several works that propose nanotargeting campaigns using the FB Custom Audience tool that requires PII data (e.g., email, mobile phone number, etc).
In the following, we discuss in detail these two groups of works. 

\subsubsection{Nanotargeting Based on non-PII Data}

Dave Kerpen \cite{kerpen} explains in a book how in 2009 \GramR{he ran an experiment attempting to reach his wife using a Facebook ad}. To this end, he configured a campaign with the following parameters \textit{<31-year-old, married, female, employees of Likeable Media, living in New York City>}. The ad included the message “I love you and miss you Carrie. Be home from Texas soon”. The ad reached Kerpen's wife.

In 2010, Korolova et al. \cite{korolova2010} leveraged the FB Ads Manager to nanotarget two specific individuals. First, they picked a friend and used the gender (female), the workplace, and the college attended by this person to configure a Facebook ad campaign to target her. The authors knew beforehand these parameters could only identify the person they wanted to target because she was the only person in her workplace that have attended the referred college. Also, they knew the targeted user had introduced the gender, college, and work information in her FB profile. As it was expected, the targeted ad was exclusively delivered to the referred friend. They repeated the experiment using a second individual, but this time they obtained information from his public FB profile. In particular, they launched a campaign using his gender, age, location, and some interests. Again, they succeeded in delivering the ad only to the targeted user. The final goal of this paper was not nanotargeting the users but showing that if you can uniquely identify an individual using a particular audience configuration, you can use the FB advertising ecosystem to unveil other personal information from that user. For instance, they revealed the age of the lady they targeted in the first experiment. To obtain the age, they extended the original audience definition by adding an age value. They launched multiple campaigns using in each of them a different age value. Among these campaigns only one delivered impressions (the remaining campaigns did not deliver a single impression).  The age used in this campaign revealed the age of the targeted user, which was indeed validated as the actual age of the lady in her FB profile. In summary, this work disclosed a privacy vulnerability of the FB advertising ecosystem. According to the authors, FB updated its advertising platform and did not allow it to run campaigns for which the actual audience size was lower than 20. Our results suggest that this limit is not currently in place since we have been able to nanotarget users on FB.

While these preliminary works offer examples on the possibility of nanotarget users on FB with non-PII data, they just present ad-hoc experiments that can be considered anecdotal pieces of evidence. Instead, our work provides the first systematic formulation that provides clear and specific guidelines to perform successful nanotargeting attacks.

%While these preliminary works offer an experiment on the possibility of nanotarget users on FB with non-PII data, our work
%provides some significant improvements and findings compared to them: $(i)$ We present the first systematic analysis on the probability to perform a nanotargeting ad campaign on FB with non-PII data. In particular,  we developed a  data-driven theoretical formulation of the number of interests that uniquely identifies a user, providing for first-time general guidelines on the ways (least popular and random interests selections) and required number of interests to successfully perform a nanotargeting attack; $(ii)$ While the cited works perform a handpicked selection of interests that guarantee beforehand and with high probability the uniqueness of the targeted user, we made a random selection of interests from the targeted users. In this sense, the experiment conducted by previous works would be more similar to the case of least popular interests selection strategy discussed in this paper; $(iii)$ The type of targeting information used in previous papers includes socio-demographic characteristics of the user such as location, attended colleague, marital status, etc that by definition narrows down significantly the size of the associated audience. Instead, there was no previous knowledge regarding how interests impact users' uniqueness. 

\subsubsection{Nanotargeting Based on PII Data}
\label{subsec:RW_nanotargeting_PII}

In addition to the regular FB Ads Manager dashboard, FB offers advertisers an alternative tool to configure ad campaigns referred to as Custom Audiences \cite{custom_audiences_fb}. As we described in Section \ref{sec:background}, an advertiser can define a Custom Audience using a list of PII items such as emails, mobile phones, etc. Facebook uses that list to identify users that are registered in the platform using the provided PII item. Advertisers can create Custom Audiences by combining the list of PII with other personalization parameters. For instance, they can create an audience including the users in the PII list that are male, or the users in the PII list that are interested in soccer. The Custom Audience feature has been used multiple times to create nanotargeting campaigns. Facebook tried to increase the privacy guarantees for users by establishing a minimum size of 20 verified users in a Custom Audience list. This limit was later increased up to 100, which is the current limit. None of these limits were enough to avoid nanotargeting.

We found several examples in the literature where the Custom Audience tool is combined with other targeting criteria to display an ad exclusively to an individual \cite{prank}\cite{harf_2017}\cite{haskins_2018}\cite{tim_shipman_2018}\cite{hawkins_2019}\cite{faddoul2019sniper}. For instance, in \cite{harf_2017} the goal was targeting a specific male user. To this end, the authors used a list of emails exclusively belonging to women except one that belonged to the targeted male user. They configured a Custom Audience that aimed to target males within the provided list of users. By doing so, it was guaranteed the ad was going to be delivered exclusively to the targeted male user. Similarly, Korolova et al. \cite{korolova2018} made use of the Custom Audience feature to attest that delivering an ad to a specific user was feasible. They bypassed the Custom Audience threshold (20 at the time this work was published) by including in the Custom Audience list 19 non-reachable FB accounts (e.g, users that have an ad blocker installed or who are not active on FB) and just 1 active account. This way, they could guarantee to deliver the ad exclusively to the targeted user. The authors proposed that the Custom Audience size limit should be increased to 1000, but as we already mentioned that limit is currently 100.

In summary, a few works demonstrate that FB Custom Audiences can be exploited to nanotarget users. However, they require to know and use PII from the targeted user. In contrast, our research aims to prove that nanotargeting can be implemented in a systematic manner using non-PII data. %\jg{FB terms of use explains that advertisers using Custom Audiences are responsible to obtain explicit permission from the users included in a Custom Audience. Failing to do so may imply the advertiser/attacker is breaking personal data regulations such as the General Data Protection Regulation (GDPR) \cite{GDPR} because they are using PII to target specific users without their explicit permission. However, FB terms of use do not require advertisers to obtain any prior permission when targeting users based on non-PII attributes such as users' interests.}  

\section{Discussion}
\label{sec:discussion}
In this section, we first discuss the potential risks of nanotargeting. Next, we describe what are the current measures FB is implementing and why they are inefficient. Finally, we propose several countermeasures that can be easily adopted by FB (and other players in the ad ecosystem) to effectively avoid nanotargeting.

\subsection{Risks Associated with Nanotargeting}

There is a body of literature referred to as \textit{psychological persuasion} that demonstrates that persuading an individual is easier if you create tailored messages to the psychological characteristics and motivations of that person\cite{pshyc_persuasion1}\cite{pshyc_persuasion2}\cite{pshyc_persuasion3}\cite{pshyc_persuasion4}\cite{pshyc_persuasion5}\cite{Kosinski_PNAS}. Some studies have demonstrated that narrowly tailored ads have much higher engaging capacity, leading, for instance, to a Click Through Rate (CTR) increase of up to 670\% \cite{mullock}. In the context of Facebook, Matz et. al \cite{Kosinski_PNAS} ran an experiment with 3.5M FB users using tailored ads together with psychological persuasion communication techniques. Authors report that they were able to increase the number of clicks up to 40\% and the purchases up to 50\% compared to non-personalized campaigns. Besides, we found a few stories on the web explaining how the FB ad ecosystem was used to persuade a specific person to perform an action. As an example, Michael Harf \cite{harf_2017} explains how he used the FB Custom Audience tool to deliver nanotargeted ads to persuade a potential client, who had previously expressed interest to move from his current digital agency, to join Harf's agency. 

There is another interesting story that, although it does not use explicitly nanotargeting as defined in our paper, is valid to illustrate some other potential risks associated with nanotargeting \cite{haskins_2018}\cite{tim_shipman_2018}. In the 2017 UK campaign, the leader of the Labour party, Jeremy Corbyn, wanted to heavily invest in digital ads encouraging voters' registration. However, the chiefs of the Labour Party campaign thought it was a bad idea. To make Corbyn happy and at the same time spend the campaign money on other objectives, the campaign chiefs invested \pounds5,000 in a Facebook campaign that exploited the Custom Audience tool to only reach Corbyn, his associates, and a few aligned journalists. By doing so, Corbyn was convinced the campaign was implemented following his instructions.

All previous \GramR{examples clearly illustrate the potential} risks of nanotargeting. First, nanotargeting can be effectively used to manipulate a user to persuade them to buy a product or to convince them to change their mind regarding a particular issue. Also, nanotargeting could be used to create a fake perception in which the user is exposed to a reality that differs from what the rest of the users see (as happened in the case of Corbyn). Finally, nanotargeting could be exploited to implement some other harmful practices such as blackmailing.

Any of the presented practices represent a very worrying manipulation of human beings. To implement such manipulation, attackers may leverage platforms like FB that allow them to deliver ads exclusively to a targeted user. This represents a privacy vulnerability for FB users that urges FB to adopt and implement efficient countermeasures.

\subsection{Current (Inefficient) Countermeasures against Nanotargeting}

The most important countermeasure Facebook implements to \GramR{prevent advertisers from targeting very narrow} audiences are the limits imposed on the minimum number of users that can form an audience. However, those limits have been proven to be completely ineffective. On the one hand, Korolova et. al \cite{korolova2010} state that, motivated by the results of their paper, Facebook disallowed configuring audiences of size smaller than 20 using the Ads Campaign Manager. Our research shows that this limit is not currently being applied. On the other hand, \GramR{FB enforces a minimum Custom Audience size of 100 users}. As presented in Section \ref{subsec:RW_nanotargeting_PII}, several works in the literature showed different ways to overcome this limit and implement nanotargeting ad campaigns using Custom Audiences.

%. On the other hand, FB establishes a minimum size of 100 users to define Custom Audiences. The works introduced in Section \ref{subsec:RW_nanotargeting_PII} have proven multiple ways to nanotarget individuals using Custom Audiences.

It is relevant to mention that, in the configuration process of one out of the 21 audiences used in our nanotargeting experiment, Facebook warned that our audience was too narrow and recommended we enlarge it to run the associated campaign.  However, we just had to substitute one interest in the list and the warning disappeared. Indeed, the referred campaign succeeded in nanotargeting the associated user. We have searched on the Facebook public documentation and we could not find any officially specified limit for the minimum audience size associated with an ad campaign.

Finally, it is also worth mentioning that a few days after our nanotargeting experiment ended, Facebook closed the account we used to run the ad campaigns. We used that same account in other research works that intensively queried the FB Ads Manager API in the recent past as well. Facebook did not provide us any explanation about the reasons leading to the removal of the account, thus we cannot confirm if it was due to the nanotargeting experiment or not. %To the best of our knowledge, our campaigns were not violating Facebook's terms of use.

Even if we assume that the account was removed due to our nanotargeting experiment, it only occurred days after the last campaign had finished. This would represent a reactive measure, which is inefficient since it did not preclude us from successfully running our nanotargeted campaigns and reaching the requested users multiple times. %As a final remark, remember that FB charged us more than 300€ from the advertising campaigns run from this account. This represents an economic benefit that they removed because they supposedly identified some inappropriate activity. 

%actual reason to remove the account was that it ran many ad campaigns to very narrow audiences, FB closed the account two weeks after we launched 12 of the 21 nanotargeting campaigns. This means, it is a reactive measure, which is inefficient since it was we managed to successfully reach the targeted users. Even more, Facebook charged us more than 300€ for an activity they may have considered suspicious. This means they obtain a benefit out of our activity. 

\subsection{Efficient Countermeasures against Nanotargeting}

Nanotargeting based on non-PII users' interests could be avoided by implementing a very simple update in the FB Ad Platform. Facebook should reduce the maximum number of interests allowed in the definition of an audience from the current limit (25) to less than 9. Our analysis in Section \ref{sec:uniqueness} indicates that this would dramatically reduce the possibility of effectively running nanotargeting ad campaigns. Besides, 
%we have contacted two experts from the digital {marketing industry who have confirmed that, based on their experience, the} fraction of ad campaigns using audiences configured with 9 or more interests in online advertising (in general) and FB (in particular) is marginal. 
\CamRd{we have consulted TAPTAP Digital \cite{taptap}, a company running Sonata \cite{sonata}, a mid-side DSP. They confirmed that it is extremely strange to configure targeted audiences with +9 interests. Indeed, <1\% of the campaigns may have such a configuration.}\footnote{\CamR{An experiment to prove that this statement is correct requires getting access to the configuration of hundreds or thousands of ad campaigns. Ad campaign configuration is a very sensitive asset for advertisers, marketing agencies, and other AdTech players, and thus we could not get access to that information. In this context, we could at least obtain an informal confirmation from two AdTech experts.}} This suggests that the proposed countermeasure is expected to have a very limited impact on FB's revenue.

The proposed measure is effective in protecting users from nanotargeting based on interests, however, it does not work to prevent PII-based nanotargeting implemented through the Facebook Custom Audience tool. Hence, to this matter, we propose a second (simple) measure that would avoid any type of nanotargeting. FB should not allow running any ad campaign whose targeted audience size is below a given limit of active users. It is important to remark that only active users (e.g., in the last month) should count for computing the audience size. 

The referred limit should not be lower than 100 and our recommendation is to set it equal to 1000. This solution would invalidate tricks like the one cited before in which a Custom Audience was integrated by only one man, and the advertising campaign was configured to target men within the Custom Audience list. If our solution was in place, FB would identify that the active audience size for such campaign is actually one and, as a result, the campaign would not be accepted. 

In summary, very basic solutions as the ones described above would provide strong protection against nanotargeting practices.

%The proposed solutions protect users from being nanotargeted and complicate the implementation of potential manipulation attempts.

%can simply reduce the maximum number of interests that can be included in a campaign from 25 to less than 9. Based on our results, this would dramatically reduce the possibility of effectively running nanotargeting ad campaigns. Also, we have contacted two industry experts in digital marketing and according to their opinion, the number of ad campaigns including more than 9 interests on FB is marginal. That means the proposed countermeasure will very likely have a negligible impact on the FB revenue.

%we have shown that most nanotargeting campaigns use PII-information through the FB Custom Audience tool. Hence, we propose another measure to proactively prevent all nanotargeting campaigns irrespectively how they define an audience. Facebook should not allow running ad campaigns for which the active audience size, i.e., users who are active on FB, is lower than a certain limit. That limit should not be lower than 100, and our recommendation is to be established in 1000. For instance, this would invalidate the case in which a custom audience list includes all women but one, and the advertisers configure an ad campaign to reach males in the referred list. The actual active audience size for that campaigns is one and if our proposal is implemented would not be accepted.

\section{Conclusion}

This paper presents two fundamental contributions. First, we provide an analytical methodology to study the number of non-PII items, i.e., interests, that make a user unique in a user base including 1.5B individuals registered on FB. This is the first analysis of user's uniqueness in a user base of a worldwide population scale. Our results indicate that the 4 rarest FB interests of a user make them unique in the mentioned user base with a 90\% probability. If we instead consider a random selection of interests, then, 22 interests would be required to make a user unique with a 90\% probability.

Second, since users' interests are actionable on FB to configure targeted ad campaigns, we leverage the results from our analysis of user uniqueness to perform real experiments in the last quarter of 2020 to implement nanotargeting ad campaigns on FB, i.e., campaigns that exclusively reach the targeted user. Our experiments prove it is possible to systematically nanotarget a user on FB based on their interests. We also propose measures to prevent potentially harmful nanotargeting attacks exploiting the FB advertising platform. 

Finally, it is worth noting that our work has only revealed the tip of the iceberg regarding how non-PII data can be used for nanotargeting purposes. Our work exclusively relies on users' interests, but an advertiser can use other available socio-demographic parameters to configure audiences in the FB Ads Manager such as the home location (country, city, zip code, etc.), workplace, college, number of children, mobile device used (iOS, Android), etc, to rapidly narrow down the audience size to nanotarget a user.  Hence, the combination of socio-demographic parameters with interests may imply that the number of non-PII items required to successfully implement a nanotargeting attack is lower than what we have reported in this paper. We plan to address this issue in our future work. We want to study the uniqueness of users as well as the probability of conducting successful nanotargeting attacks on FB when considering a combination of socio-demographic parameters and interests in the configuration of audiences.

% \begin{acks}
%-------------------------------------------------------------------------------
 \section*{Acknowledgments}
%-------------------------------------------------------------------------------
% Here we will acknowledge the funding that lead to these results.
\CamR{This research received funding from the European Union’s Horizon 2020 innovation action programme under the PIMCITY project (Grant 871370) and the TESTABLE project (Grant 101019206); the Ministerio de Economía, Industria y Competitividad, Spain, and the European Social Fund(EU), under the Ramón y Cajal programme (Grant RyC-2015-17732); the Ministerio de Educación, Cultura y Deporte, Spain, through the FPU programme (Grant FPU16/05852); the Agencia Estatal de Investigación (AEI) under the ACHILLES project (Grant PID2019-104207RB-I00/AEI/10.13039/501100011033); the Community of Madrid synergic project EMPATIA-CM (Grant Y2018/TCS-5046); the Fundación BBVA under the project AERIS; and the Vienna Science and Technology Fund through the project ``Emotional Well-Being in the Digital Society'' (Grant VRG16-005).}
% \end{acks}

\balance
\bibliographystyle{ACM-Reference-Format}
\bibliography{references} 


\begin{thebibliography}{38}

%%% the Web address.

\ifx \showCODEN    \undefined \def \showCODEN     #1{\unskip}     \fi
\ifx \showDOI      \undefined \def \showDOI       #1{#1}\fi
\ifx \showISBNx    \undefined \def \showISBNx     #1{\unskip}     \fi
\ifx \showISBNxiii \undefined \def \showISBNxiii  #1{\unskip}     \fi
\ifx \showISSN     \undefined \def \showISSN      #1{\unskip}     \fi
\ifx \showLCCN     \undefined \def \showLCCN      #1{\unskip}     \fi
\ifx \shownote     \undefined \def \shownote      #1{#1}          \fi
\ifx \showarticletitle \undefined \def \showarticletitle #1{#1}   \fi
\ifx \showURL      \undefined \def \showURL       {\relax}        \fi
% invisible to TeX
\providecommand\bibfield[2]{#2}
\providecommand\bibinfo[2]{#2}
\providecommand\natexlab[1]{#1}
\providecommand\showeprint[2][]{arXiv:#2}

\bibitem[\protect\citeauthoryear{Bennett and Lanning}{Bennett and
  Lanning}{2007}]%
        {Netflix_prize}
\bibfield{author}{\bibinfo{person}{J. Bennett} {and} \bibinfo{person}{S.
  Lanning}.} \bibinfo{year}{2007}\natexlab{}.
\newblock \showarticletitle{The Netflix Prize}. In
  \bibinfo{booktitle}{\emph{Proceedings of the KDD Cup Workshop 2007}}.
  \bibinfo{publisher}{ACM}, \bibinfo{address}{New York}, \bibinfo{pages}{3--6}.
\newblock
\urldef\tempurl%
\url{http://www.cs.uic.edu/~liub/KDD-cup-2007/NetflixPrize-description.pdf}
\showURL{%
\tempurl}


\bibitem[\protect\citeauthoryear{Cesario, Higgins, and Scholer}{Cesario
  et~al\mbox{.}}{2008}]%
        {pshyc_persuasion1}
\bibfield{author}{\bibinfo{person}{Joseph Cesario}, \bibinfo{person}{E.~Tory
  Higgins}, {and} \bibinfo{person}{Abigail~A. Scholer}.}
  \bibinfo{year}{2008}\natexlab{}.
\newblock \showarticletitle{{Regulatory Fit and Persuasion: Basic Principles
  and Remaining Questions}}.
\newblock \bibinfo{journal}{\emph{Social and Personality Psychology Compass}}
  \bibinfo{volume}{2}, \bibinfo{number}{1} (\bibinfo{year}{2008}),
  \bibinfo{pages}{444--463}.
\newblock
\urldef\tempurl%
\url{https://doi.org/10.1111/j.1751-9004.2007.00055.x}
\showDOI{\tempurl}
\showeprint{https://onlinelibrary.wiley.com/doi/pdf/10.1111/j.1751-9004.2007.00055.x}


\bibitem[\protect\citeauthoryear{de~Montjoye, Hidalgo, Verleysen, and
  Blondel}{de~Montjoye et~al\mbox{.}}{2013}]%
        {de2013unique}
\bibfield{author}{\bibinfo{person}{Yves-Alexandre de Montjoye},
  \bibinfo{person}{César~A Hidalgo}, \bibinfo{person}{Michel Verleysen}, {and}
  \bibinfo{person}{Vincent~D Blondel}.} \bibinfo{year}{2013}\natexlab{}.
\newblock \showarticletitle{{Unique in the Crowd: The privacy bounds of human
  mobility}}.
\newblock \bibinfo{journal}{\emph{Scientific reports}} \bibinfo{volume}{3},
  \bibinfo{number}{1} (\bibinfo{year}{2013}), \bibinfo{pages}{1376}.
\newblock
\showISSN{2045-2322}
\urldef\tempurl%
\url{https://www.nature.com/articles/srep01376}
\showURL{%
\tempurl}


\bibitem[\protect\citeauthoryear{De~Montjoye, Radaelli, Singh,
  et~al\mbox{.}}{De~Montjoye et~al\mbox{.}}{2015}]%
        {de2015unique}
\bibfield{author}{\bibinfo{person}{Yves-Alexandre De~Montjoye},
  \bibinfo{person}{Laura Radaelli}, \bibinfo{person}{Vivek~Kumar Singh},
  {et~al\mbox{.}}} \bibinfo{year}{2015}\natexlab{}.
\newblock \showarticletitle{{Unique in the shopping mall: On the
  reidentifiability of credit card metadata}}.
\newblock \bibinfo{journal}{\emph{Science}} \bibinfo{volume}{347},
  \bibinfo{number}{6221} (\bibinfo{year}{2015}), \bibinfo{pages}{536--539}.
\newblock
\urldef\tempurl%
\url{https://science.sciencemag.org/content/347/6221/536}
\showURL{%
\tempurl}


\bibitem[\protect\citeauthoryear{Dubois, Rucker, and Galinsky}{Dubois
  et~al\mbox{.}}{2016}]%
        {pshyc_persuasion5}
\bibfield{author}{\bibinfo{person}{David Dubois}, \bibinfo{person}{Derek~D.
  Rucker}, {and} \bibinfo{person}{Adam~D. Galinsky}.}
  \bibinfo{year}{2016}\natexlab{}.
\newblock \showarticletitle{{Dynamics of Communicator and Audience Power: The
  Persuasiveness of Competence versus Warmth}}.
\newblock \bibinfo{journal}{\emph{Journal of Consumer Research}}
  \bibinfo{volume}{43}, \bibinfo{number}{1} (\bibinfo{date}{Feb.}
  \bibinfo{year}{2016}), \bibinfo{pages}{68--85}.
\newblock
\showISSN{0093-5301}
\urldef\tempurl%
\url{https://doi.org/10.1093/jcr/ucw006}
\showDOI{\tempurl}
\showeprint{https://academic.oup.com/jcr/article-pdf/43/1/68/7049938/ucw006.pdf}


\bibitem[\protect\citeauthoryear{Erikson and Erikson}{Erikson and
  Erikson}{1998}]%
        {erikson1998life}
\bibfield{author}{\bibinfo{person}{Erik~H Erikson} {and}
  \bibinfo{person}{Joan~M Erikson}.} \bibinfo{year}{1998}\natexlab{}.
\newblock \bibinfo{booktitle}{\emph{The life cycle completed (extended
  version)}}.
\newblock \bibinfo{publisher}{WW Norton \& Company}.
\newblock


\bibitem[\protect\citeauthoryear{EU}{EU}{2016}]%
        {GDPR}
\bibfield{author}{\bibinfo{person}{EU}.} \bibinfo{year}{2016}\natexlab{}.
\newblock \bibinfo{title}{{Regulation (EU) 2016/679 of the European Parliament
  and of the Council of 27 April 2016 on the protection of natural persons with
  regard to the processing of personal data and on the free movement of such
  data, and repealing Directive 95/46/EC (General Data Protection
  Regulation)}}.
\newblock \bibinfo{howpublished}{European Union}.
\newblock
\urldef\tempurl%
\url{http://eur-lex.europa.eu/eli/reg/2016/679/oj}
\showURL{%
\tempurl}
\newblock
\shownote{accessed on 21 September, 2021.}


\bibitem[\protect\citeauthoryear{Faddoul, Kapuria, and Lin}{Faddoul
  et~al\mbox{.}}{2019}]%
        {faddoul2019sniper}
\bibfield{author}{\bibinfo{person}{Marc Faddoul}, \bibinfo{person}{Rohan
  Kapuria}, {and} \bibinfo{person}{Lily Lin}.} \bibinfo{year}{2019}\natexlab{}.
\newblock \showarticletitle{SNIPER AD TARGETING}.
\newblock \bibinfo{journal}{\emph{Berkeley School of Information}}
  (\bibinfo{date}{May} \bibinfo{year}{2019}).
\newblock
\urldef\tempurl%
\url{https://www.ischool.berkeley.edu/projects/2019/sniper-ad-targeting}
\showURL{%
\tempurl}


\bibitem[\protect\citeauthoryear{Faizullabhoy and Korolova}{Faizullabhoy and
  Korolova}{2018}]%
        {korolova2018}
\bibfield{author}{\bibinfo{person}{Irfan Faizullabhoy} {and}
  \bibinfo{person}{Aleksandra Korolova}.} \bibinfo{year}{2018}\natexlab{}.
\newblock \showarticletitle{{Facebook's Advertising Platform: New Attack
  Vectors and the Need for Interventions}}. In
  \bibinfo{booktitle}{\emph{Workshop on Technology and Consumer Protection
  (ConPro 2018)}}.
\newblock
\showeprint[arxiv]{1803.10099}~[cs.CY]
\urldef\tempurl%
\url{https://arxiv.org/abs/1803.10099}
\showURL{%
\tempurl}


\bibitem[\protect\citeauthoryear{FB}{FB}{2017}]%
        {fb_end_report_2016}
\bibfield{author}{\bibinfo{person}{FB}.} \bibinfo{year}{2017}\natexlab{}.
\newblock \bibinfo{title}{{Facebook Reports Fourth Quarter and Full Year 2016
  Results}}.
\newblock \bibinfo{howpublished}{Facebook Inc.}.
\newblock
\urldef\tempurl%
\url{https://investor.fb.com/investor-news/press-release-details/2017/Facebook-Reports-Fourth-Quarter-and-Full-Year-2016-Results/default.aspx}
\showURL{%
\tempurl}
\newblock
\shownote{accessed on 21 September, 2021.}


\bibitem[\protect\citeauthoryear{FB}{FB}{2021a}]%
        {custom_audiences_fb}
\bibfield{author}{\bibinfo{person}{FB}.} \bibinfo{year}{2021}\natexlab{a}.
\newblock \bibinfo{title}{{About Custom Audiences}}.
\newblock \bibinfo{howpublished}{Facebook Inc.}.
\newblock
\urldef\tempurl%
\url{https://www.facebook.com/business/help/744354708981227}
\showURL{%
\tempurl}
\newblock
\shownote{accessed 11 July, 2021.}


\bibitem[\protect\citeauthoryear{FB}{FB}{2021b}]%
        {FBadsprefs}
\bibfield{author}{\bibinfo{person}{FB}.} \bibinfo{year}{2021}\natexlab{b}.
\newblock \bibinfo{title}{{Facebook Ad Preferences}}.
\newblock \bibinfo{howpublished}{Facebook Inc.}.
\newblock
\urldef\tempurl%
\url{https://www.facebook.com/adpreferences/}
\showURL{%
\tempurl}
\newblock
\shownote{accessed on 21 September, 2021.}


\bibitem[\protect\citeauthoryear{FB}{FB}{2021c}]%
        {FBadsmanager}
\bibfield{author}{\bibinfo{person}{FB}.} \bibinfo{year}{2021}\natexlab{c}.
\newblock \bibinfo{title}{{Facebook Ads Manager}}.
\newblock \bibinfo{howpublished}{Facebook Inc.}.
\newblock
\urldef\tempurl%
\url{https://www.facebook.com/ads/manager}
\showURL{%
\tempurl}
\newblock
\shownote{accessed on 21 September, 2021.}


\bibitem[\protect\citeauthoryear{FB}{FB}{2021d}]%
        {fb_q4_2020}
\bibfield{author}{\bibinfo{person}{FB}.} \bibinfo{year}{2021}\natexlab{d}.
\newblock \bibinfo{title}{{Facebook Reports Fourth Quarter and Full Year 2020
  Results}}.
\newblock \bibinfo{howpublished}{Facebook Inc.}.
\newblock
\urldef\tempurl%
\url{https://investor.fb.com/investor-news/press-release-details/2021/Facebook-Reports-Fourth-Quarter-and-Full-Year-2020-Results/default.aspx}
\showURL{%
\tempurl}
\newblock
\shownote{accessed on 21 September, 2021.}


\bibitem[\protect\citeauthoryear{FDVT}{FDVT}{2016a}]%
        {fdvt_policy_privacy}
\bibfield{author}{\bibinfo{person}{FDVT}.} \bibinfo{year}{2016}\natexlab{a}.
\newblock \bibinfo{title}{{FDVT: Privacy Agreement}}.
\newblock
\newblock
\urldef\tempurl%
\url{https://www.fdvt.org/privacy_agreement.html}
\showURL{%
\tempurl}
\newblock
\shownote{accessed on 21 September, 2021.}


\bibitem[\protect\citeauthoryear{FDVT}{FDVT}{2016b}]%
        {fdvt_terms_of_use}
\bibfield{author}{\bibinfo{person}{FDVT}.} \bibinfo{year}{2016}\natexlab{b}.
\newblock \bibinfo{title}{{FDVT: Terms of Use}}.
\newblock
\newblock
\urldef\tempurl%
\url{https://www.fdvt.org/terms_of_use}
\showURL{%
\tempurl}
\newblock
\shownote{accessed on 21 September, 2021.}


\bibitem[\protect\citeauthoryear{FDVT}{FDVT}{2021}]%
        {fdvt_url}
\bibfield{author}{\bibinfo{person}{FDVT}.} \bibinfo{year}{2021}\natexlab{}.
\newblock \bibinfo{title}{{FDVT: Data Valuation Tool for Facebook™ Users
  Website}}.
\newblock
\newblock
\urldef\tempurl%
\url{https://fdvt.org/}
\showURL{%
\tempurl}
\newblock
\shownote{accessed on 21 September, 2021.}


\bibitem[\protect\citeauthoryear{Gendronneau, Yıldız, Hsiao, Stepanek, Abel,
  Hoorens, Wiśniowski, Zagheni, Fiorio, and Weber}{Gendronneau
  et~al\mbox{.}}{2019}]%
        {fbto100}
\bibfield{author}{\bibinfo{person}{Cloé Gendronneau}, \bibinfo{person}{Dilek
  Yıldız}, \bibinfo{person}{Yuan Hsiao}, \bibinfo{person}{Martin Stepanek},
  \bibinfo{person}{Guy Abel}, \bibinfo{person}{Stijn Hoorens},
  \bibinfo{person}{Arkadiusz Wiśniowski}, \bibinfo{person}{Emilio Zagheni},
  \bibinfo{person}{Lee Fiorio}, {and} \bibinfo{person}{Ingmar Weber}.}
  \bibinfo{year}{2019}\natexlab{}.
\newblock \bibinfo{title}{{Measuring Labour Mobility and Migration Using Big
  Data - Exploring the potential of social-media data for measuring EU mobility
  flows and stocks of EU movers}}.
\newblock
\newblock
\urldef\tempurl%
\url{https://ec.europa.eu/social/BlobServlet?docId=22084}
\showURL{%
\tempurl}


\bibitem[\protect\citeauthoryear{Golle}{Golle}{2006}]%
        {Golle:2006:RUS:1179601.1179615}
\bibfield{author}{\bibinfo{person}{Philippe Golle}.}
  \bibinfo{year}{2006}\natexlab{}.
\newblock \showarticletitle{{Revisiting the Uniqueness of Simple Demographics
  in the US Population}}. In \bibinfo{booktitle}{\emph{Proceedings of the 5th
  ACM Workshop on Privacy in Electronic Society}} (Alexandria, Virginia, USA)
  \emph{(\bibinfo{series}{WPES '06})}. \bibinfo{publisher}{ACM},
  \bibinfo{address}{New York, NY, USA}, \bibinfo{pages}{77--80}.
\newblock
\showISBNx{1-59593-556-8}
\urldef\tempurl%
\url{https://doi.org/10.1145/1179601.1179615}
\showDOI{\tempurl}


\bibitem[\protect\citeauthoryear{Gonz\'{a}lez Caba\~{n}as, Cuevas, and
  Cuevas}{Gonz\'{a}lez Caba\~{n}as et~al\mbox{.}}{2017}]%
        {FDVT}
\bibfield{author}{\bibinfo{person}{Jos\'{e} Gonz\'{a}lez Caba\~{n}as},
  \bibinfo{person}{\'{A}ngel Cuevas}, {and} \bibinfo{person}{Rub\'{e}n
  Cuevas}.} \bibinfo{year}{2017}\natexlab{}.
\newblock \bibinfo{booktitle}{\emph{FDVT: Data Valuation Tool for Facebook
  Users}}.
\newblock \bibinfo{publisher}{Association for Computing Machinery},
  \bibinfo{address}{New York, NY, USA}, \bibinfo{pages}{3799–3809}.
\newblock
\showISBNx{9781450346559}
\urldef\tempurl%
\url{https://doi.org/10.1145/3025453.3025903}
\showURL{%
\tempurl}


\bibitem[\protect\citeauthoryear{Harf}{Harf}{2017}]%
        {harf_2017}
\bibfield{author}{\bibinfo{person}{Michael Harf}.}
  \bibinfo{year}{2017}\natexlab{}.
\newblock \bibinfo{title}{{Sniper Targeting on Facebook: How to Target ONE
  specific person with super targeted ads}}.
\newblock \bibinfo{howpublished}{Medium}.
\newblock
\urldef\tempurl%
\url{https://medium.com/@MichaelH_3009/sniper-targeting-on-facebook-how-to-target-one-specific-person-with-super-targeted-ads-515ba6e068f6}
\showURL{%
\tempurl}
\newblock
\shownote{accessed on 21 September, 2021.}


\bibitem[\protect\citeauthoryear{Haskins}{Haskins}{2018}]%
        {haskins_2018}
\bibfield{author}{\bibinfo{person}{Caroline Haskins}.}
  \bibinfo{year}{2018}\natexlab{}.
\newblock \bibinfo{title}{Facebook ad micro-targeting can manipulate individual
  politicians}.
\newblock \bibinfo{howpublished}{The Outline}.
\newblock
\urldef\tempurl%
\url{https://theoutline.com/post/5411/facebook-ad-micro-targeting-can-manipulate-individual-politicians}
\showURL{%
\tempurl}
\newblock
\shownote{accessed on 21 September, 2021.}


\bibitem[\protect\citeauthoryear{Hawkins}{Hawkins}{2019}]%
        {hawkins_2019}
\bibfield{author}{\bibinfo{person}{Jonathan Hawkins}.}
  \bibinfo{year}{2019}\natexlab{}.
\newblock \bibinfo{title}{{Facebook Ads Sniper Method: How to Put Your Ad in
  front of ONE Specific Person}}.
\newblock \bibinfo{howpublished}{Jonathan Hawkins}.
\newblock
\urldef\tempurl%
\url{https://jonathanhawkinsofficial.com/blog/facebook-ads-sniper-method-how-to-put-your-ad-in-front-of-one-specific-person}
\showURL{%
\tempurl}
\newblock
\shownote{accessed on 21 September, 2021.}


\bibitem[\protect\citeauthoryear{Hirsh, Kang, and Bodenhausen}{Hirsh
  et~al\mbox{.}}{2012}]%
        {pshyc_persuasion4}
\bibfield{author}{\bibinfo{person}{Jacob~B. Hirsh}, \bibinfo{person}{Sonia~K.
  Kang}, {and} \bibinfo{person}{Galen~V. Bodenhausen}.}
  \bibinfo{year}{2012}\natexlab{}.
\newblock \showarticletitle{{Personalized Persuasion: Tailoring Persuasive
  Appeals to Recipients’ Personality Traits}}.
\newblock \bibinfo{journal}{\emph{Psychological Science}} \bibinfo{volume}{23},
  \bibinfo{number}{6} (\bibinfo{year}{2012}), \bibinfo{pages}{578--581}.
\newblock
\urldef\tempurl%
\url{https://doi.org/10.1177/0956797611436349}
\showDOI{\tempurl}
\showeprint{https://doi.org/10.1177/0956797611436349}
\newblock
\shownote{PMID: 22547658.}


\bibitem[\protect\citeauthoryear{IMDb}{IMDb}{2021}]%
        {IMDb}
\bibfield{author}{\bibinfo{person}{IMDb}.} \bibinfo{year}{2021}\natexlab{}.
\newblock \bibinfo{title}{{The Internet Movie Database}}.
\newblock
\newblock
\urldef\tempurl%
\url{https://www.imdb.com/}
\showURL{%
\tempurl}
\newblock
\shownote{accessed on 21 September, 2021.}


\bibitem[\protect\citeauthoryear{Kerpen}{Kerpen}{2011}]%
        {kerpen}
\bibfield{author}{\bibinfo{person}{Dave Kerpen}.}
  \bibinfo{year}{2011}\natexlab{}.
\newblock \bibinfo{booktitle}{\emph{Likeable social media : how to delight your
  customers, create an irresistible brand, and be generally amazing on Facebook
  (and other social networks)}}.
\newblock \bibinfo{publisher}{McGraw-Hill}.
\newblock
\showISBNx{9780071813723}
\showLCCN{2011008764}


\bibitem[\protect\citeauthoryear{Korolova}{Korolova}{2010}]%
        {korolova2010}
\bibfield{author}{\bibinfo{person}{Aleksandra Korolova}.}
  \bibinfo{year}{2010}\natexlab{}.
\newblock \showarticletitle{{Privacy Violations Using Microtargeted Ads: A Case
  Study}}. In \bibinfo{booktitle}{\emph{{ICDMW} 2010, The 10th {IEEE}
  International Conference on Data Mining Workshops}},
  \bibfield{editor}{\bibinfo{person}{Wei Fan}, \bibinfo{person}{Wynne Hsu},
  \bibinfo{person}{Geoffrey~I. Webb}, \bibinfo{person}{Bing Liu},
  \bibinfo{person}{Chengqi Zhang}, \bibinfo{person}{Dimitrios Gunopulos}, {and}
  \bibinfo{person}{Xindong Wu}} (Eds.). \bibinfo{publisher}{{IEEE} Computer
  Society}, \bibinfo{address}{Sydney, Australia}, \bibinfo{pages}{474--482}.
\newblock
\urldef\tempurl%
\url{https://doi.org/10.1109/ICDMW.2010.137}
\showDOI{\tempurl}


\bibitem[\protect\citeauthoryear{Matz, Kosinski, Nave, and Stillwell}{Matz
  et~al\mbox{.}}{2017}]%
        {Kosinski_PNAS}
\bibfield{author}{\bibinfo{person}{S.~C. Matz}, \bibinfo{person}{M. Kosinski},
  \bibinfo{person}{G. Nave}, {and} \bibinfo{person}{D.~J. Stillwell}.}
  \bibinfo{year}{2017}\natexlab{}.
\newblock \showarticletitle{Psychological targeting as an effective approach to
  digital mass persuasion}.
\newblock \bibinfo{journal}{\emph{Proceedings of the National Academy of
  Sciences}} \bibinfo{volume}{114}, \bibinfo{number}{48}
  (\bibinfo{year}{2017}), \bibinfo{pages}{12714--12719}.
\newblock
\showISSN{0027-8424}
\urldef\tempurl%
\url{https://doi.org/10.1073/pnas.1710966114}
\showDOI{\tempurl}
\showeprint{https://www.pnas.org/content/114/48/12714.full.pdf}


\bibitem[\protect\citeauthoryear{Moon}{Moon}{2002}]%
        {pshyc_persuasion3}
\bibfield{author}{\bibinfo{person}{Youngme Moon}.}
  \bibinfo{year}{2002}\natexlab{}.
\newblock \showarticletitle{{Personalization and Personality: Some Effects of
  Customizing Message Style Based on Consumer Personality}}.
\newblock \bibinfo{journal}{\emph{Journal of Consumer Psychology}}
  \bibinfo{volume}{12}, \bibinfo{number}{4} (\bibinfo{year}{2002}),
  \bibinfo{pages}{313--325}.
\newblock
\urldef\tempurl%
\url{https://doi.org/10.1016/S1057-7408(16)30083-3}
\showDOI{\tempurl}
\showeprint{https://onlinelibrary.wiley.com/doi/pdf/10.1016/S1057-7408\%2816\%2930083-3}


\bibitem[\protect\citeauthoryear{{Mullock}, {Groom}, , and {Lee}}{{Mullock}
  et~al\mbox{.}}{2010}]%
        {mullock}
\bibfield{author}{\bibinfo{person}{J. {Mullock}}, \bibinfo{person}{S. {Groom}},
  \bibinfo{person}{}, {and} \bibinfo{person}{P. {Lee}}.}
  \bibinfo{year}{2010}\natexlab{}.
\newblock \bibinfo{title}{International online behavioural advertising survey
  2010}.
\newblock \bibinfo{howpublished}{Osborne Clarke}.
\newblock


\bibitem[\protect\citeauthoryear{Narayanan and Shmatikov}{Narayanan and
  Shmatikov}{2008}]%
        {Netflix_Reidentification}
\bibfield{author}{\bibinfo{person}{Arvind Narayanan} {and}
  \bibinfo{person}{Vitaly Shmatikov}.} \bibinfo{year}{2008}\natexlab{}.
\newblock \showarticletitle{{Robust De-Anonymization of Large Sparse
  Datasets}}. In \bibinfo{booktitle}{\emph{Proceedings of the 2008 IEEE
  Symposium on Security and Privacy}} \emph{(\bibinfo{series}{SP '08})}.
  \bibinfo{publisher}{IEEE Computer Society}, \bibinfo{address}{USA},
  \bibinfo{pages}{111–125}.
\newblock
\showISBNx{9780769531687}
\urldef\tempurl%
\url{https://doi.org/10.1109/SP.2008.33}
\showDOI{\tempurl}


\bibitem[\protect\citeauthoryear{{Sonata DSP}}{{Sonata DSP}}{2021}]%
        {sonata}
\bibfield{author}{\bibinfo{person}{{Sonata DSP}}.}
  \bibinfo{year}{2021}\natexlab{}.
\newblock \bibinfo{title}{{Global Platform for Mobile-Centric Audience
  Engagement}}.
\newblock
\newblock
\urldef\tempurl%
\url{https://www.sonataplatform.com}
\showURL{%
\tempurl}
\newblock
\shownote{accessed on 21 September, 2021.}


\bibitem[\protect\citeauthoryear{Su, Shukla, Goel, and Narayanan}{Su
  et~al\mbox{.}}{2017}]%
        {su}
\bibfield{author}{\bibinfo{person}{Jessica Su}, \bibinfo{person}{Ansh Shukla},
  \bibinfo{person}{Sharad Goel}, {and} \bibinfo{person}{Arvind Narayanan}.}
  \bibinfo{year}{2017}\natexlab{}.
\newblock \showarticletitle{{De-Anonymizing Web Browsing Data with Social
  Networks}}. In \bibinfo{booktitle}{\emph{Proceedings of the 26th
  International Conference on World Wide Web}} (Perth, Australia)
  \emph{(\bibinfo{series}{WWW '17})}. \bibinfo{publisher}{International World
  Wide Web Conferences Steering Committee}, \bibinfo{address}{Republic and
  Canton of Geneva, CHE}, \bibinfo{pages}{1261–1269}.
\newblock
\showISBNx{9781450349130}
\urldef\tempurl%
\url{https://doi.org/10.1145/3038912.3052714}
\showDOI{\tempurl}


\bibitem[\protect\citeauthoryear{Sweeney}{Sweeney}{2000}]%
        {sweeney2000simple}
\bibfield{author}{\bibinfo{person}{Latanya Sweeney}.}
  \bibinfo{year}{2000}\natexlab{}.
\newblock \showarticletitle{Simple demographics often identify people
  uniquely}.
\newblock \bibinfo{journal}{\emph{Health (San Francisco)}}
  \bibinfo{volume}{671}, \bibinfo{number}{2000} (\bibinfo{year}{2000}),
  \bibinfo{pages}{1--34}.
\newblock


\bibitem[\protect\citeauthoryear{Swichkow}{Swichkow}{2014}]%
        {prank}
\bibfield{author}{\bibinfo{person}{Brian Swichkow}.}
  \bibinfo{year}{2014}\natexlab{}.
\newblock \bibinfo{title}{{The Ultimate Retaliation: Pranking My Roommate With
  Targeted Facebook Ads}}.
\newblock \bibinfo{howpublished}{Ghost Influence}.
\newblock
\urldef\tempurl%
\url{http://ghostinfluence.com/the-ultimate-retaliation-pranking-my-roommate-with-targeted-facebook-ads/}
\showURL{%
\tempurl}
\newblock
\shownote{accessed on 21 September, 2021.}


\bibitem[\protect\citeauthoryear{{TAPTAP Digital}}{{TAPTAP Digital}}{2021}]%
        {taptap}
\bibfield{author}{\bibinfo{person}{{TAPTAP Digital}}.}
  \bibinfo{year}{2021}\natexlab{}.
\newblock \bibinfo{title}{{Omnichannel advertising and marketing intelligence
  powered by location}}.
\newblock
\newblock
\urldef\tempurl%
\url{https://www.taptapdigital.com/}
\showURL{%
\tempurl}
\newblock
\shownote{accessed on 21 September, 2021.}


\bibitem[\protect\citeauthoryear{Tim~Shipman}{Tim~Shipman}{2018}]%
        {tim_shipman_2018}
\bibfield{author}{\bibinfo{person}{Political~Editor Tim~Shipman}.}
  \bibinfo{year}{2018}\natexlab{}.
\newblock \bibinfo{title}{{Labour HQ used Facebook ads to deceive Jeremy
  Corbyn}}.
\newblock \bibinfo{howpublished}{The Sunday Times}.
\newblock
\urldef\tempurl%
\url{https://www.thetimes.co.uk/article/labour-hq-used-facebook-ads-to-deceive-corbyn-3hvn0jzr8}
\showURL{%
\tempurl}
\newblock
\shownote{accessed on 21 September, 2021.}


\bibitem[\protect\citeauthoryear{Wheeler, Petty, and Bizer}{Wheeler
  et~al\mbox{.}}{2005}]%
        {pshyc_persuasion2}
\bibfield{author}{\bibinfo{person}{S. Wheeler}, \bibinfo{person}{Richard
  Petty}, {and} \bibinfo{person}{George Bizer}.}
  \bibinfo{year}{2005}\natexlab{}.
\newblock \showarticletitle{{Self-Schema Matching and Attitude Change:
  Situational and Dispositional Determinants of Message Elaboration}}.
\newblock \bibinfo{journal}{\emph{Journal of Consumer Research}}
  \bibinfo{volume}{31} (\bibinfo{date}{March} \bibinfo{year}{2005}),
  \bibinfo{pages}{787--797}.
\newblock
\urldef\tempurl%
\url{https://doi.org/10.1086/426613}
\showDOI{\tempurl}


\end{thebibliography}

\appendix
\section*{Appendix}
\label{sec:appendix}

\begin{table}[b]
\resizebox{.95\hsize}{!}{%
\begin{tabular}{rlr|rlr}
\textbf{code} & \textbf{country} & \textbf{users (M)} &\textbf{code} & \textbf{country} & \textbf{users (M)} \\
\hline
  US & United States & 203 & DZ & Algeria & 16 \\ 
  IN & India & 161 & NG & Nigeria & 16  \\
  BR & Brazil & 114 & AU & Australia & 15 \\ 
  ID & Indonesia & 91  & IQ & Iraq & 14  \\
  MX & Mexico & 70 & PL & Poland & 14 \\ 
  PH & Philippines & 56 & SA & Saudi Arabia & 14 \\
  TR & Turkey & 46 & ZA & South Africa & 14 \\ 
  TH & Thailand & 42  & MA & Morocco & 13  \\
  VN & Vietnam & 42 & VE & Venezuela & 13 \\ 
  GB & United Kingdom & 39  & CL & Chile & 12  \\
  EG & Egypt & 33 & MM & Myanmar & 12 \\ 
  FR & France & 33  & RU & Russia & 12  \\
  DE & Germany & 30 & NL & Netherlands & 10 \\
  IT & Italy & 30  & EC & Ecuador & 9.80  \\
  AR & Argentina & 29 & RO & Romania & 8.60 \\
  PK & Pakistan & 28  & AE & UA Emirates & 7.70 \\
  CO & Colombia & 26 & NP & Nepal & 6.70 \\
  JP & Japan & 26  & BE & Belgium & 6.50  \\
  BD & Bangladesh & 23 & SE & Sweden & 6.20 \\ 
  ES & Spain & 23  & TN & Tunisia & 6.10 \\
  CA & Canada & 22 & KE & Kenya & 6 \\ 
  MY & Malaysia & 20  & PT & Portugal & 5.90  \\
  PE & Peru & 19 & UA & Ukraine & 5.90 \\ 
  KR & South Korea & 18  & GT & Guatemala & 5.50  \\
  TW & Taiwan & 18 & HU & Hungary & 5.30 \\
  \hline
\end{tabular}%
}
\caption{List of the 50 countries included in our queries to the FB Ads Manager and their associated number of users in millions.}
\label{tab:list_50countries}
\end{table}

\section{FB User Base for Uniqueness Analysis}
\label{ap:tabla50}

\GramR{At the time we collected the data (January 2017) the Facebook API did not allow an audience to be defined from the entire world}. Instead, it was compulsory to insert a location or group of locations to define the geographical coverage of the defined audience. The maximum number of locations allowed in an ad campaign was 50. To maximize the user base of our uniqueness analysis on FB we selected the top 50 countries in terms of active users. \GramR{These 50 countries included 1.5B active FB users which corresponded to 81\% of the overall Facebook users \cite{fb_end_report_2016}}. Table \ref{tab:list_50countries} lists the 50 considered countries along with the number of FB users.

%Nowadays this option is available under a field called ``Worldwide - Region''. 
%Table \ref{tab:list_50countries} shows the set of countries we used to retrieve the potential reach information in our queries.  However, back in 2017, the advertiser needed to define the specific locations to target. Besides, there was a maximum number of locations allowed. The advertiser could not specify more than 50 countries in an audience. Then, to increase the number of users addressed in our research, we built a set with the 50 largest countries by the number of FB users and then perform the requests for this country setting. At that time, in January 2017, the top 50 countries on FB represented 1.5 Billion users which corresponded to 81\% of the overall Facebook users \cite{fb_end_report_2016}.

\section{Location Breakdown of Users}
\label{ap:distribution}

\jg{The only compulsory parameter to define an audience in the FB Ads Manager is a location (e.g., country, region, zip code, etc.). This means, an audience can be configured by a single location, but if you want to use multiple attributes \GramR{at least one of them must to be a location}. Based on this restriction, in the registration process of \CamR{the FDVT} browser extension users had to obligatorily fill in their location (i.e., country of residence). Otherwise, the browser extension could not retrieve any information from the FB Ads Manager API, and subsequently, could not provide users with the estimated revenue they generate for FB. Our user base of 2,390 users was distributed across 80 different locations. Table \ref{tab:user_locs} shows the number of users per country.}

\begin{table}[!b]
%\vspace{1.5cm}
\resizebox{.95\hsize}{!}{%
\begin{tabular}{rlr|rlr}
\textbf{code} & \textbf{country} & \textbf{users} & \textbf{code} & \textbf{country} & \textbf{users} \\
  \hline
  ES & Spain & 1131 & AU & Australia &   2 \\ 
  FR & France & 335 & CY & Cyprus &   2 \\ 
  MX & Mexico & 122 & DO & Dominican Republic &   2 \\ 
  AR & Argentina & 115 & GR & Greece &   2 \\ 
  EC & Ecuador &  89 & HK & Hong Kong SAR China &   2 \\ 
  PE & Peru &  78 & ID & Indonesia &   2 \\ 
  CA & Canada &  61 & IE & Ireland &   2 \\ 
  CO & Colombia &  48 & LU & Luxembourg &   2 \\ 
  US & United States &  40 & PL & Poland &   2 \\ 
  BE & Belgium &  36 & RE & Réunion &   2 \\ 
  UY & Uruguay &  35 & AL & Albania &   1 \\ 
  GB & United Kingdom &  26 & AM & Armenia &   1 \\ 
  CH & Switzerland &  24 & AO & Angola &   1 \\ 
  PT & Portugal &  21 & AX & Åland Islands &   1 \\ 
  VE & Venezuela &  18 & BG & Bulgaria &   1 \\ 
  SV & El Salvador &  17 & BT & Bhutan &   1 \\ 
  CL & Chile &  14 & CI & Côte d’Ivoire &   1 \\ 
  PY & Paraguay &  13 & CR & Costa Rica &   1 \\ 
  DE & Germany &  11 & CZ & Czechia &   1 \\ 
  IT & Italy &  11 & DJ & Djibouti &   1 \\ 
  BO & Bolivia &   9 & GI & Gibraltar &   1 \\ 
  MA & Morocco &   8 & GN & Guinea &   1 \\ 
  BR & Brazil &   6 & IN & India &   1 \\ 
  GT & Guatemala &   6 & IQ & Iraq &   1 \\ 
  HN & Honduras &   6 & LK & Sri Lanka &   1 \\ 
  NI & Nicaragua &   6 & LT & Lithuania &   1 \\ 
  NL & Netherlands &   6 & MG & Madagascar &   1 \\ 
  PA & Panama &   6 & MO & Macao SAR China &   1 \\ 
  TN & Tunisia &   6 & MU & Mauritius &   1 \\ 
  BD & Bangladesh &   5 & NC & New Caledonia &   1 \\ 
  SE & Sweden &   4 & NP & Nepal &   1 \\ 
  TH & Thailand &   4 & NZ & New Zealand &   1 \\ 
  AD & Andorra &   3 & PH & Philippines &   1 \\ 
  AT & Austria &   3 & PM & St. Pierre \& Miquelon &   1 \\ 
  DK & Denmark &   3 & PR & Puerto Rico &   1 \\ 
  DZ & Algeria &   3 & RO & Romania &   1 \\ 
  FI & Finland &   3 & RS & Serbia &   1 \\ 
  PK & Pakistan &   3 & RU & Russia &   1 \\ 
  SN & Senegal &   3 & RW & Rwanda &   1 \\ 
  AF & Afghanistan &   2 & TW & Taiwan &   1 \\ 
  \hline
\end{tabular}%
}
\caption{Complete breakdown of the number of users per location in our 2,390 users' dataset retrieved from \CamR{the FDVT} browser extension.}
%\vspace{1cm}
\label{tab:user_locs}
\end{table}
%\section{Facebook Interests Popularity}
%\label{ap:section_cdf}

%The 25th-75th interquartile interval of the distribution is [113,193, 1,719,925] users with a median audience size of 418,530 users. 

%To understand the popularity of these interests, we extracted the audience size reported by the FB Marketing API for each of them. Out of the 1.5M interests included in our dataset, 98,982 are unique. Each of the interests has an audience size (number of Facebook users labeled with that interest in their profiles) associated. As shown in Figure \ref{fig:cdf_interests}, we studied the information reported by FB on the number of users included in each of the 98,982 unique interests of our dataset. We observe that these interests cover a wide range of popularity and thus are representative to conduct our analysis. 

\section{Demographic Analysis}
\label{ap:demographic_analysis}

An intriguing question is if the number of interests that make a user unique on FB shows significant differences across different demographic groups. To answer this question, we obtain the value of $N(LP)_{0.9}$ and $N(R)_{0.9}$ across three demographic parameters: gender, age, and location.  This demographic analysis just aims to illustrate that there may be differences in nanotargeting users according to demographic parameters.

We have selected $P$ = 0.9 for two reasons: (i) it reveals the number of interests that uniquely identifies a user on FB with a very high probability (0.9); (ii) $N(LP)_{0.9}$ and $N(R)_{0.9}$ are both below 25 (the maximum number of interests that can be used to define an audience on FB) and thus are actionable in practice to perform nanotargeting. 

\subsection{Gender analysis}

\begin{figure*}[t]
\centering
\begin{minipage}[t]{0.315\textwidth}
	\includegraphics[width=1\columnwidth]{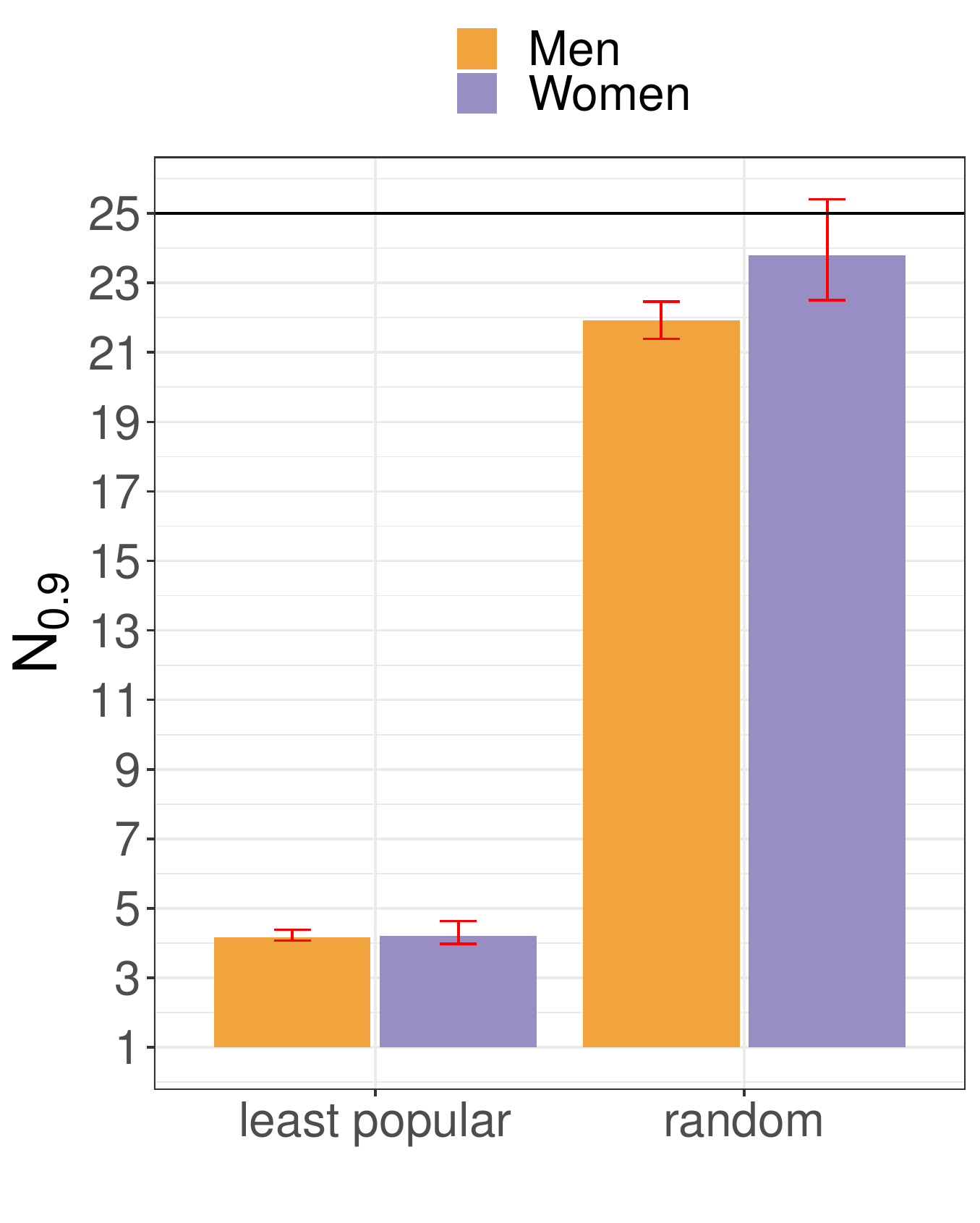}\hfill
	%\vspace{-0.5cm}
	\caption{Uniqueness analysis across gender. $N(LP)_{0.9}$ (left) and $N(R)_{0.9}$ (right) for men (yellow) and women (purple). The figure includes the 95\% confidence interval of the results.}
	\label{fig:genders}
\end{minipage}
\hfill
\begin{minipage}[t]{0.315\textwidth}
	\includegraphics[width=1\columnwidth]{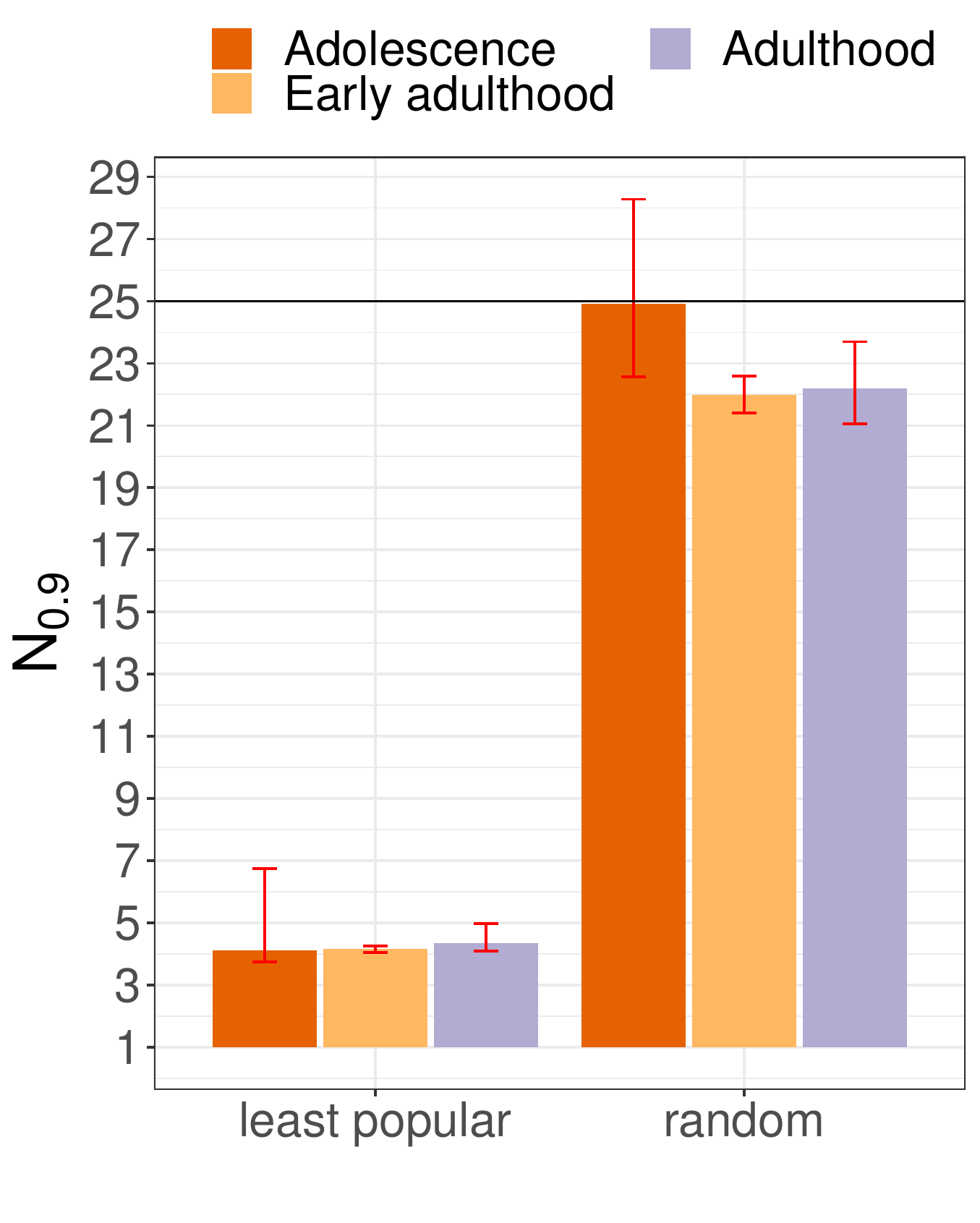}\hfill
	%\vspace{-0.5cm}
	\caption{Uniqueness analysis across age groups. $N(LP)_{0.9}$ (left) and $N(R)_{0.9}$ (right) for adolescence (orange), early adulthood (yellow) and adulthood (purple) groups. The figure includes the 95\% confidence interval of the results.}
	\label{fig:ages}
\end{minipage}
\hfill
\begin{minipage}[t]{0.315\textwidth}
	\includegraphics[width=\hsize]{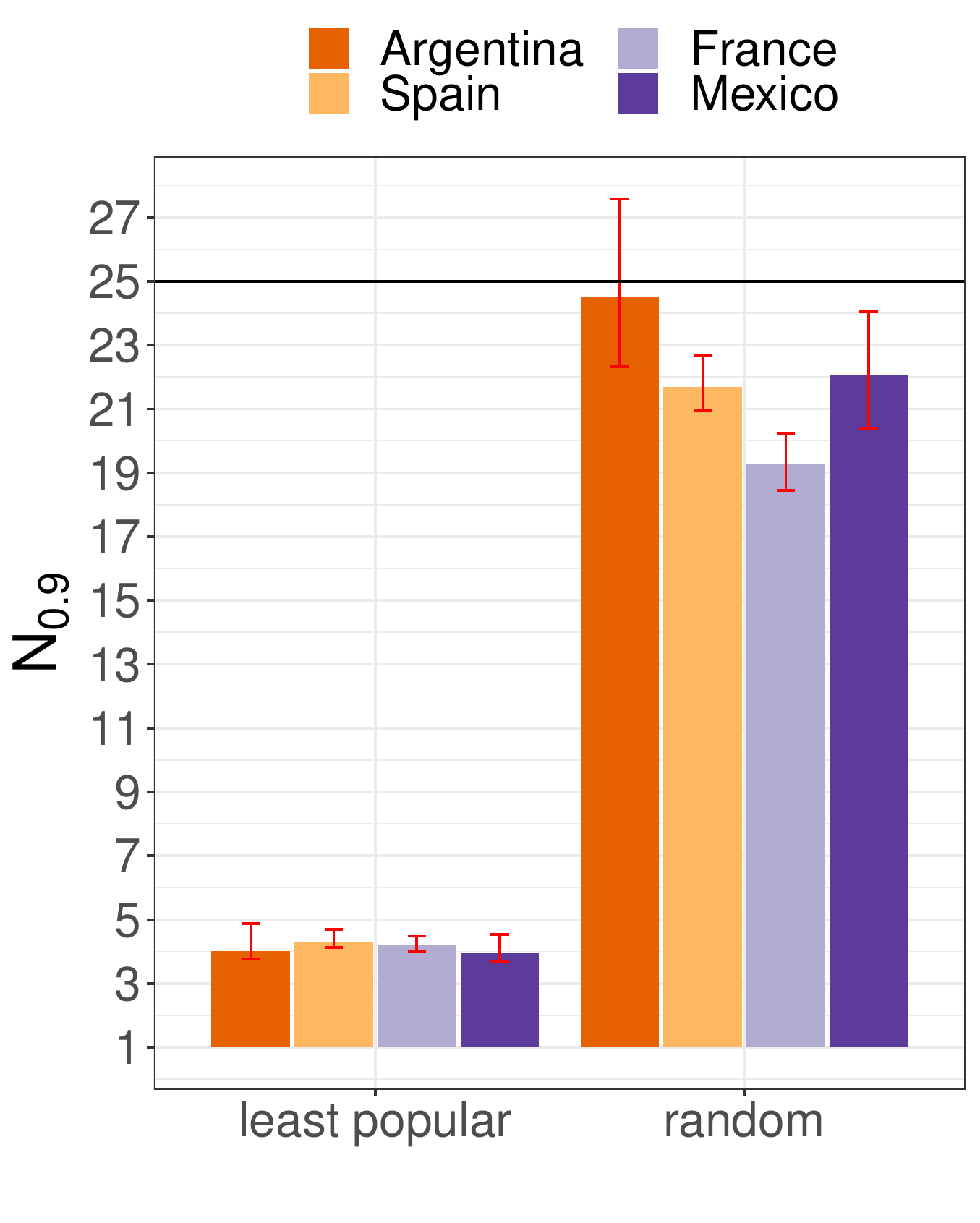}\hfill
	%\vspace{-0.5cm}
	\caption{Uniqueness analysis across countries. $N(LP)_{0.9}$ (left) and $N(R)_{0.9}$ (right) for Argentina (orange), Spain (yellow), France (light purple) and Mexico (dark purple). The figure includes the 95\% confidence interval of the results.}
	\label{fig:countries}
 \end{minipage}
\end{figure*}

We divide our dataset into men (1,949 users) and women (347 users) and compute $N(LP)_{0.9}$ and $N(R)_{0.9}$ for each group. Figure \ref{fig:genders} shows the result in the form of a bar plot. Note that we present the 95\% confidence interval of our fitting model in the form of an error bar in each bar plot. 

We observe that $N(LP)_{0.9}$ is almost the same for men (4.16) and women (4.20), indicating that the number of interests that make a man or a woman unique within a worldwide population-scale user base is similar and close to 4. 

$N(R)_{0.9}$ presents a larger difference, being 23.80 for women and 21.92 for men. This finding indicates that an attacker would need to infer (roughly) two interests more to nanotarget a woman in comparison to a man. This suggests that women's interest profiles are slightly more private than men's and thus are harder to nanotarget.

%separately for men and women as a bar plot the value of $N_{0.95}$ along with the 95\% confidence interval values derived from our fitting model. The figure depicts the results for the case of selecting the least popular interests and the case of selecting the interests at random.

%$N_{0.9}$ is almost the same in the case of men (4.16) and women (4.20) for the case of selecting the least popular interests. Contrarily, when we focus on random interests, we observe a relevant difference in gender since $N_{0.9}=23.80$ for women and $N_{0.9}=21.92$ for men. That means women on FB require two more interests than men to be unequivocally identified on FB with a 0.9 probability. 

%In conclusion, according to our results, women's interests' profile is slightly more private than men's interests' profile. In practice, that means, that is slightly easier to nanotarget a man than a woman. 

\subsection{Age analysis.}

We divided the users in our dataset into the following age groups based on the division proposed by Erikson et al. \cite{erikson1998life}: 13-19 (Adolescence), 20-39 (Early-Adulthood), 40-64 (Adulthood), and 65+ (Maturity). The number of users in the Adolescence, Early-Adulthood, Adulthood, and Maturity groups are 117; 1,374; 578; and 19, respectively. Due to the low number of users forming the Maturity group we decided to exclude it from the analysis. 

Figure \ref{fig:ages} shows the value of $N(LP)_{0.9}$ and $N(R)_{0.9}$ for the Adolescence, Early-Adulthood, and Adulthood age groups along with the 95 confidence interval of our model.

The values of $N(LP)_{0.9}$ are very similar in all considered age groups (4.11, 4.16, and 4.45 for the Adolescence, Early-Adulthood, and Adulthood groups, respectively). This result indicates that the uniqueness of a user in FB seems not to be correlated with their age group. 

If we focus now on the $N(R)_{0.9}$ values, we observe that users in the Early-Adulthood and Adulthood can be nanotargeted with a 90\% success probability with 22 interests ($N(R)_{0.9}$ = 21.99 and 22.20 for Early-Adulthood and Adulthood, respectively). Nanotargeting users in the Adolescence group for the same probability is harder since it requires 25 interest ($N(R)_{0.9}$ = 24.92).

%When considering the least popular interests, users in all the three age groups roughly require 4 interests to be unequivocally identified on FB (4.11, 4.16, and 4.45 for the Adolescence, Early-Adulthood, and Adulthood groups, respectively). Similarly, in the case of using random interests, users in the Early-Adulthood ($N_{0.9}$ = 21.99) and Adulthood ($N_{0.9}$ = 22.20) groups would require $\sim22$ interests to be unequivocally identified. However, users in the Adolescence group  (($N_{0.9}$ = 24.92) would require 3 interests more. 

%In summary, our results suggest that nanotargeting young users ($<20$ years) may be more difficult than nanotargeting users ranging between 20 and 65 years old. 

\subsection{Location analysis.}

While our dataset includes users from 80 different countries (see Table \ref{tab:user_locs}), most of them present a low number of users. Therefore, to derive meaningful results, we select those countries for which we have more than 100 users in our dataset. These are: Spain (1131 users), France (335), Mexico (122), and Argentina (115). 

Following the same analysis we conducted for gender and age groups, Figure  \ref{fig:countries} shows bar plots capturing the values of $N(LP)_{0.9}$ and $N(R)_{0.9}$ for the considered countries along with the 95\% confidence intervals provided by the fitting model in the form or error bars. 

As in the case of gender and age, $N_{0.9}(LP)$ is very similar for the four considered countries (3.96, 4.03, 4.21, and 4.29 for Mexico, Argentina, France, and Spain, respectively), confirming that none of the considered demographic parameters seem to be relevant to impact the user uniqueness on FB. 

$N_{0.9}(R)$ values are 19.28, 21.7, 22.05, and 24.49 for France, Spain, Mexico, and Argentina, respectively. This indicates that conducting a nanotargeted ad campaign would be notably easier in France compared to Argentina since an attacker would need to infer 5 interests less in the former country to perform a nanotargeted campaign to a user with a success probability of 90\%. This result suggests that the location is a factor that may be relevant in the number of interests required to nanotarget a user on FB.

\emph{In summary, our demographic analysis reveals that women, adolescents, and users from Argentina (compared to France, Spain, and Mexico) are better protected from nanotargeting attacks based on random interest selection.}

\begin{figure}[t]
\centering
	\includegraphics[width=.97\hsize]{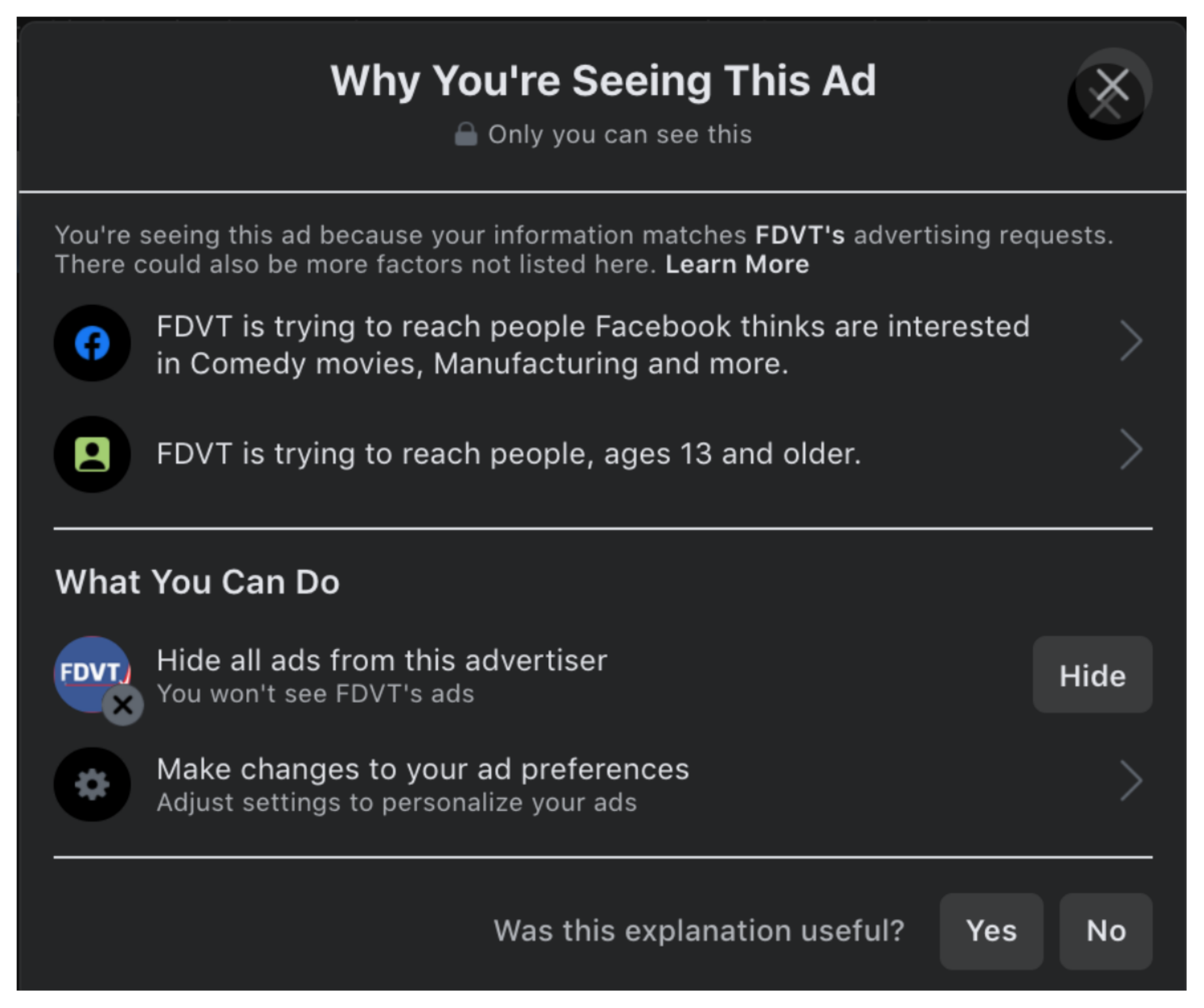}\hfill
	\caption{Snapshot of the \emph{``Why am I seeing this ad''} window associated to the ad impression of the campaign targeting user 3 with 12 interests.}% For the double-blind review process, we have removed the name of our web browser extension from the figure wherever it appeared.}
	\label{fig:ad_ap2}
  \hfill
\end{figure}
\begin{figure}[t]
\centering
	\includegraphics[width=1\hsize]{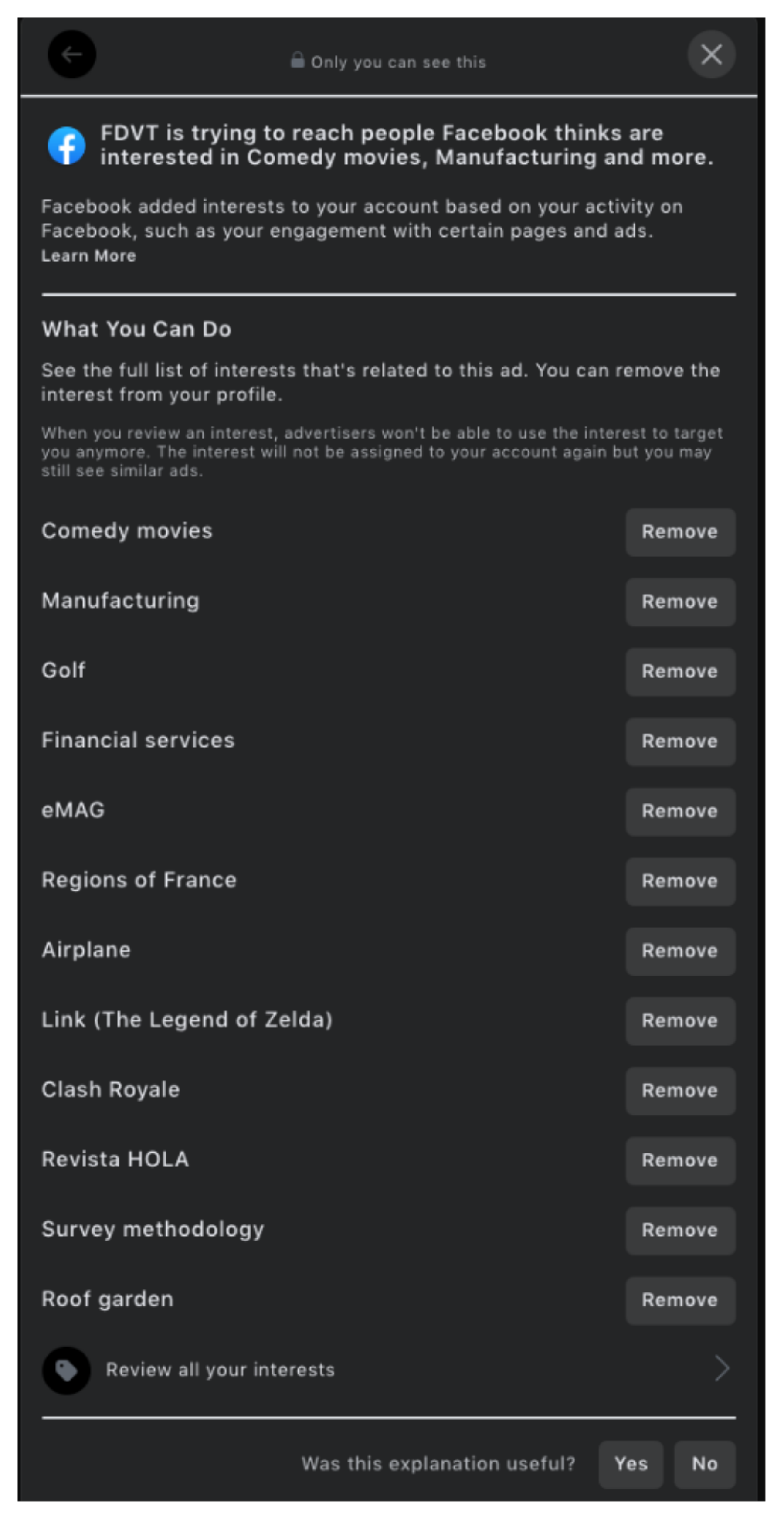}\hfill
	%\vspace{0.1cm}
	\caption{Snapshot of the list of interests used in the ad campaign targeting User 3 with 12 interests obtained from the \emph{``Why am I seeing this ad''} function.}% For the double-blind review process, we have removed the name of our web browser extension from the figure wherever it appeared.}
	\label{fig:ad_ap3}
  \hfill
\end{figure}

\section{Nanotargeted Ad Impressions Proofs}
\label{ap:ads_fdvt}

Each nanotargeting campaign in our experiment used one exclusive ad creativity, which included a text to identify the targeted user and the number of interests used in the associated campaign. %This text is included in the bottom right corner of the creativity. Figure \ref{fig:ad_ap1} illustrates the case of the ad creativity used in the campaign targeting User 3 with 12 interests.  
Using this information, the targeted authors could easily identify the ads associated with the experiment when they saw them in their FB newsfeed.

Besides, each ad creativity is linked to a different landing page hosted on our web server. The authors were instructed to click on the nanotargeted ads whenever they see them on the FB newsfeed. Those clicks were logged in our web server along with their timestamp.

Finally, users were asked to capture several snapshots associated with the received ads from the nanotargeting campaigns:

\begin{enumerate}

\item The three authors had to capture an image of the ad impression received in their FB newsfeed. Figure \ref{fig:ad_ap1} illustrates the ad received by User 3 associated with the nanotargeting campaign configured with 12 random interests.

\item The authors had to obtain a snapshot of the field \emph{``Why am I seeing this ad''} where Facebook shows the user what are the reasons for receiving the associated ad. Figure \ref{fig:ad_ap2} shows a snapshot of the \emph{``Why am I seeing this ad''} reasons associated with the ad impression of the campaign targeting User 3 with 12 interests. 

\item The authors were also asked to click on the option offered in the \emph{``Why am I seeing this ad''} window that provides the specific targeting attributes used in the ad campaign associated with the ad. This allows retrieving the actual list of interests used in the ad campaign associated with the ad. Figure \ref{fig:ad_ap3} shows a snapshot of the list of interests included in the \emph{``Why am I seeing this ad''} field for the ad campaign targeting User 3 with 12 random interests.

\end{enumerate}

\end{document}